\documentclass[12pt]{article}
\usepackage{a41} %
\usepackage{epsfig} %
\usepackage{amssymb} %
\usepackage{amsmath} %
\usepackage{verbatim} %
\usepackage{cite} %
\newcommand{\Mvec}{\mbox{\rm\bf M}}

\newcommand{\beq}{\begin{equation}}
\newcommand{\eeq}{\end{equation}}
\newcommand{\bea}{\begin{eqnarray}}
\newcommand{\eea}{\end{eqnarray}}

\newcommand{\GeV}{${\rm GeV}$}

\newcommand{\gsim}{\raisebox{-0.07cm}{$\, \stackrel{>}{{\scriptstyle
\sim}}\, $}}
\newcommand{\EM}{{\rm E-}}

\newcommand\MeV{\,\mbox{MeV}}

\newcommand\ONE{\mbox{\boldmath $1$}}

\newcommand\Lvec{\mbox{\boldmath $L$}}

\newcommand\Pvec{\mbox{\boldmath $P$}}
\newcommand\Uvec{\mbox{\boldmath $U$}}

\newcommand{\nn}{\nonumber}
\newcommand{\tx}{\textrm}
\newcommand{\spar}{{\stackrel{\rightarrow}{\Rightarrow}}}
\newcommand{\sant}{{\stackrel{\rightarrow}{\Leftarrow}}}

\newcounter{lin}

\graphicspath{/usr1/hboett/plots/qcd_pol/}
\graphicspath{./}

\begin{document}
\begin{titlepage}

\begin{flushleft}
DESY 09--131 \hfill {\tt arXiv:1005.3113 [hep-ph]} 
\\
SFB-CPP/10--032\\
May 2010 \\
\end{flushleft}

\vspace{3cm}
\noindent
\begin{center}
{\LARGE\bf QCD Analysis of Polarized Deep Inelastic}\\

\vspace*{2mm}
\noindent
{\LARGE\bf Scattering Data }
\end{center}
\begin{center}

\vspace{4cm}
{\large Johannes Bl\"umlein and Helmut B\"ottcher}

\vspace{2cm}
{\it 
Deutsches Elektronen Synchrotron, DESY}\\

\vspace{3mm}
{\it  Platanenallee 6, D--15738 Zeuthen, Germany}\\
\vspace{3cm}
\end{center}
\begin{abstract}
\noindent
A QCD analysis of the world data on polarized deep inelastic scattering is presented 
in next--to--leading order, including the heavy flavor Wilson coefficient in leading
order in the fixed flavor number scheme. New parameterizations are derived for the 
quark and gluon distributions and the value of $\alpha_s(M_z^2)$ is determined. The 
impact of the variation of both the renormalization and factorization scales on the 
distributions and the value of $\alpha_s$ is studied. We obtain $\alpha_s^{\rm NLO}(M_Z^2)  
= 0.1132~~\begin{array}{l}  + 0.0056 \\ -0.0095 \end{array}$. The first moments of the 
polarized twist--2 parton distribution functions are calculated with correlated errors 
to allow for comparisons with results from lattice QCD simulations. Potential higher twist 
contributions to the structure function $g_1(x,Q^2)$ are determined both for proton and 
deuteron targets.
\end{abstract}

\end{titlepage}

\newpage
\sloppy

\section{Introduction}
\label{sec:intro}

\vspace{2mm}
\noindent
The short--distance structure of the nucleon spin still consists a developing field. Nucleons 
as composite fermions obtain their spin in terms of  a superposition of the spins and orbital
angular momenta of their constituents, the quarks and gluons. It came as a surprise when the 
European Muon Collaboration (EMC) published its result~\cite{EMCp} more than 20 years ago,
which showed that the quarks do contribute only by a small fraction to the nucleon's
spin. The obvious conclusion was to assume that the spin of the gluons and the orbital angular 
momenta of all constituents have to account for the missing fraction. This result initiated 
activities worldwide both on the experimental and the theoretical side in order
to understand this spin puzzle and, finally, the spin structure of the nucleon.   

Experiments performed at CERN, SLAC, DESY and JLAB \cite{EMCp,E142n,HERMn,E154QCD,E154n,SMCpd,
E143pd,HERMpd,E155d,E155p,COMPd,JLABn,CLA1pd,CLA2pd,COMP1} have contributed a vast amount of 
experimental data on inclusive polarized deeply inelastic lepton--nucleon scattering (DIS) during 
the last years. The main interest in measuring the short distance structure of 
polarized nucleons has somewhat moved from determining the first moments of the twist--2 parton 
distributions to the extraction of their $x$-dependence in the measured region and their scaling 
violations due to QCD--evolution. 
At large enough four-momentum transfer $Q^2 = -q^2$, the structure function $g_1(x,Q^2)$ mainly 
receives twist--2 contributions\footnote{Twist--3 contributions are connected by target mass 
effects, cf.~\cite{BT}.} and is related to the polarized twist--2 parton distributions. 

In the present paper a QCD analysis of the polarized deep--inelastic world data is performed
at next-to-leading order (NLO). Due to a larger set of new data which has become available 
recently the present analysis extends and updates 
earlier 
investigations~\cite{ABFR,AAC,GRSV,Bourrely:2005kw,LSS,LSS_HT,BB,AK,DSSV,SMCQCD,E154QCD,E155p,DelDebbio:2009sq}.
In the QCD-fit we determine the 
flavor singlet and non-singlet contributions of the polarized parton densities together with 
the QCD-scale $\Lambda_{\rm QCD}$ with correlated errors. The measurement of the strong coupling 
constant $\alpha_s(M_Z^2)$ from polarized deep-inelastic data does not reach the same precision
as in the unpolarized case \cite{BBG,MSTW,JR,ABKM} since the measurement is based on an 
asymmetry and
the present analysis is performed in NLO. However, a consistent analysis requires the determination
of the QCD-scale $\Lambda_{\rm QCD}$ along with the parameters of the non-perturbative input 
distributions. Also, it is interesting to see which value of $\alpha_s(M_Z^2)$ is obtained in 
comparison to other deep--inelastic analyses.
At a given scale $Q^2$ the Mellin moments
of the parton distribution functions can be calculated under some assumptions on their 
extrapolation outside the measured region towards small and very large values of the Bjorken 
variable $x$. We also analyze, to which extend the present data contain higher twist contributions.

The paper is organized as follows. In Section~2 the basic formalism is lined out. The data analysis
is described in Section~3. In Section~4 details of the fit are given and Section~5 deals with 
the error analysis. The results of the QCD analysis are presented in Section~6. In Section~7 we
determine potential higher twist contributions and Section~8 contains the conclusions.
In the Appendix we describe the {\tt FORTRAN}-code through which the polarized parton distributions and
structure functions can be obtained for numerical analyses.
\section{Basic Formalism}
\label{sec:qcdana}

\vspace{2mm}
\noindent
The twist--2 contributions to the spin--dependent structure function
${g}_1(x,Q^2)$ are given in terms of a {Mellin} convolution of
the polarized singlet $\Delta \Sigma$, the gluon $\Delta G$ and the
flavor non-singlet (NS) $\Delta q_i^{\rm NS}$ densities with the corresponding
Wilson coefficient functions $\Delta C_i^A$ by  
\bea
\label{eqg1}
\hspace*{-1mm}
{g}_1(x,Q^2) &=& \!\frac{1}{2} \sum_{i=1}^{N_f} e_i^2 
\int_x^1 \frac{dz}{z} 
                     \Bigg [ \frac{ 1}{ N_f}\, \Delta \Sigma 
\left(\frac{x }{ z},\mu_f^2\right) \Delta C_q^S \left(z,\frac{Q^2 }
{\mu_f^2}\right) + \Delta G\left(\frac{x }{ z},\mu_f^2\right)
\Delta C_G\left(z,\frac{Q^2}{ \mu_f^2}, \frac{m_c^2}{Q^2}\right) \nn\\
                      & & \hspace*{2.3cm}
+ \Delta q_i^{NS}\left(\frac{x }{
z},\mu_f^2\right) \Delta C_q^{NS}\left(z,\frac{Q^2 }{ \mu_f^2}\right)
\Bigg ]~. 
\eea
Here $x$ is the Bjorken variable, $e_i$ denotes the charge of the $i$th quark 
flavor in units of the elementary charge and $N_f$ is the number of light flavors.   
The scale $\mu_f$ denotes the factorization scale which is introduced
to absorb the collinear singularities into the renormalized partonic distribution
functions. In addition to the factorization scale there is the
renormalization scale $\mu_r$ of the strong coupling constant
$\alpha_s(\mu_r^2)$. The gluonic Wilson coefficient $\Delta C_G$ accounts
for the massless as well the massive contributions due to charm quark
production for $W^2 > (2 m_c + m_N)^2$ 
with $m_c = 1.5~{\rm GeV}$, \cite{HEAV1}, at first order.~\footnote{2nd order corrections were 
calculated in the asymptotic range $Q^2 \gg m^2$ in Refs.~\cite{HQ2}.}
For the implementation of the Wilson coefficients in Mellin space we refer to
\cite{AB1}. 
The parton densities and the Wilson coefficient
functions are dependent on these scales and obey corresponding
renormalization group equations, while the structure function
${g}_1(x,Q^2)$, as a physical observable, is independent of the
choice of both scales $\mu_f^2$ and $\mu_r^2$.  

The polarized singlet and non--singlet parton densities which occur in 
Eq.~(\ref{eqg1}) are expressed by the individual quark flavor
contributions as 
\bea
\label{eqfS}
\Delta \Sigma\left(z,\mu_f^2\right) &=& \sum_{i=1}^{N_f}
\Big[\Delta q_i\left(z,\mu_f^2\right)+
\Delta {\bar{q_i}}\left(z,\mu_f^2\right)\Big]~,
\\
\label{eqfNS}
\Delta q_i^{NS}\left(z,\mu_f^2\right) &=& 
\Delta q_i\left(z,\mu_f^2\right) + \Delta
{\bar{q_i}}\left(z,\mu_f^2\right) 
-\frac{1 }{ N_f} \Delta \Sigma\left(z,\mu_f^2\right)~,
\eea
where $\Delta q_i$ denotes the polarized quark distribution of the
$i$th light flavor. We will consider three light quark flavors in the present
analysis and treat charm quark production through the gluon fusion process in 
leading order \cite{HEAV1} in the fixed flavor number scheme for partons. As well
known, the following flavor non--singlet combinations contribute in the case of
pure photon exchange considering proton and neutron targets, 
\bea
\Delta_p^{\rm NS} + \Delta_n^{\rm NS} &=& \frac{1}{9}\left[ 
  (\Delta u + \Delta \bar{u})
+ (\Delta d + \Delta \bar{d}) \right] - \frac{2}{9} (\Delta s + \Delta \bar{s})
\\
\Delta_p^{\rm NS} - \Delta_n^{\rm NS} &=& \frac{1}{3}\left[ 
  (\Delta u + \Delta \bar{u})
- (\Delta d + \Delta \bar{d}) \right]~. 
\eea
As usual one assumes 
\bea
\frac{1}{6} \Delta q_{\rm sea} = \Delta \bar{q}_i~,~~~i = u,d,s,
\eea
although later highest precision measurements could even reveal a breaking of this relation.
The distribution  
\bea
\Delta \bar{d} - \Delta \bar{u}
\eea
cannot be measured in DIS data, whereas polarized Drell--Yan data are well suited for 
its determination, cf. e.g. \cite{ABKM}. For the measurement of the polarized strangeness distribution 
function one would wish to have  sufficiently precise polarized di--muon samples available.
Nonetheless first flavor separations have been attempted in performing SIDIS analyses, 
cf.~\cite{DSSV}. 

The running coupling constant $a_s = \alpha_s/(4\pi)$ is obtained as the solution of 
\bea
\label{eqarun}
\frac{d a_s(\mu_r^2)}{d \log(\mu_r^2)}  = - \beta_0 a_s^2(\mu_r^2)
- \beta_1 a_s^3( \mu_r^2) + O(a_s^4)~,
\eea
where, in the $\overline{\rm MS}$--scheme, the coefficients of the
$\beta$--function are given by 
\bea
\label{eqbeta}
\beta_0 &=&  \frac{11}{3} C_A - \frac{4}{3} T_F N_f~, \nonumber\\
\beta_1 &=&  \frac{34}{3} C_A^2 - \frac{20}{3} C_A T_F N_f -
4 C_F T_F N_f~,
\eea
with the color factors $C_A = 3$, $T_F = 1/2$, and $C_F =4/3$. The matching of the scale 
$\Lambda_{\rm QCD}^{ N_f}$ is performed at  $Q^2 = m_c^2,~m_b^2$, 
with $m_c = 1.5$~GeV and $m_b = 4.5$~GeV.

In the present  analysis the spin--dependent structure functions 
${g}_1^p(x,Q^2)$ and ${g}_1^n(x,Q^2)$ will be considered referring to
$N_f=3$ light partonic flavors, i.e. $i = u,d,s$. The
spin--dependent structure function ${g}_1^d(x,Q^2)$ is represented in terms of
${g}_1^p(x,Q^2)$ and ${g}_1^n(x,Q^2)$ using the relation   
\beq
{g}_1^{d}(x, Q^2)=\frac{1}{ 2 }\left(1 -\frac{3}{ 2} \omega_D\right)
\left[{g}_1^{p}(x, Q^2)+{g}_1^{n}(x, Q^2)\right]~,
\eeq
where $\omega_D = 0.05 \pm 0.01$ is the $D$-state wave probability for the
deuteron \cite{OMEGD}.

The change of the parton densities with respect to the factorization 
scale $\mu_f^2 = Q^2$ is described by the evolution equations, which
read 
\bea
\label{eqNSE}
\frac{\partial \Delta q_i^{\rm NS}(x,Q^2)}{\partial \log Q^2} &=& 
\Delta P_{qq}^{NS}(x,a_s) \otimes \Delta
q_i^{\rm NS}(x,Q^2)\\
\label{eqSIE}
\frac{\partial}{\partial \log Q^2} \left(\begin{array}{c} 
\Delta \Sigma(x,Q^2) \\ \Delta G(x,Q^2) \end{array} \right) &=& 
\Delta \Pvec(x,a_s) \otimes 
\left(\begin{array}{c} 
\Delta \Sigma(x,Q^2) \\ \Delta G(x,Q^2) \end{array} \right), 
\eea
with
\bea
\label{eqSIP}
\Delta \Pvec(x,a_s) &\equiv&  \left(\begin{array}{cc}
\Delta P_{qq}(x,a_s) & 2N_f \Delta P_{qg}(x,a_s) \\ \Delta P_{gq}(x,a_s) &
\Delta P_{gg}(x,a_s ) 
\end{array} \right)~.
\eea
The symbol $\otimes$ denotes the {Mellin} convolution
\bea
\label{eqmeco}
[A \otimes B](x) = \int_0^1 dx_1 dx_2 \delta(x-x_1 x_2)
A(x_1) B(x_2)~.
\eea
The spin--dependent coefficient functions $\Delta C_i^A$ and anomalous
dimensions $\Delta P_{ij}$ are calculated to next-to-leading order in the
$\overline{\rm MS}$--scheme \cite{SPLIT,FP,COEF}, which we use in the present analysis.
As seen from Eqs. (\ref{eqNSE}) and (\ref{eqSIE}), the flavor non-singlet densities
$\Delta q_i^{NS}$ evolve independently, while
$\Delta \Sigma$ and $\Delta G$ are coupled in the evolution. 

In order to solve the evolution equations,
a {Mellin} transformation of Eqs.~(\ref{eqNSE}, \ref{eqSIE}) and 
the polarized parton  densities $\Delta f(x)$ is
being performed by calculating its $N$th {Mellin} moment~:
\bea
\label{eqMEL}
\Mvec[\Delta f](N) = \int_0^1 dx~x^{N-1} \Delta f(x)~, N \geq N_0, N \in {\bf R}~.
\eea
Here $N_0$ is chosen such that the integral (\ref{eqMEL}) converges. 
Under this transformation the {Mellin} convolution $\otimes$ turns
into an ordinary product. After the transformation has been 
performed the argument $N$ is analytically continued to the complex
plane. This also requires analytic continuations of harmonic sums \cite{HSUM}, which
is outlined in Refs.~\cite{ANCONT} in detail. 
The fundamental method of solving Eqs.~(\ref{eqNSE}) and
(\ref{eqSIE}) is described in the literature in detail, see
e.g. Refs.~\cite{FP,GRV,BV}.  

To next-to-leading order (NLO) the solution of the flavor non--singlet and 
singlet evolution equations are given by
\bea
\label{eqNSS}
\Delta q^{\rm NS}_i(N,a_s) &=& \left(\frac{a_s}{a_0}\right)
^{-P_{\rm NS}^{(0)}/\beta_0} \left[1 - \frac{1}{\beta_0}(a_s-a_0)
\left(\Delta P_{\rm NS}^{(1)} - \frac{\beta_1}{\beta_0} P_{\rm NS}^{(0)}\right)
\right] \Delta q^{\rm NS}_i(N,a_0)~,
\\
\label{eqSIS}
\left(\begin{array}{c} \Delta \Sigma(N,a_s) \\ \Delta G(N,a_s)
\end{array} \right)
&=& \left[ \ONE + a_s \Uvec_1(N)\right] \Lvec(N,a_s,a_0) 
\left[\ONE - a_0 \Uvec_1(N)\right]
\left(\begin{array}{c} \Delta \Sigma(N,a_0) \\ \Delta G(N,a_0)
\end{array} \right)~.
\eea
Here $P_{\rm NS}^{(0)}$ is the leading order splitting function for the quark-quark transition. 
$\Delta P_{\rm NS}^{(1)}$ denotes the NLO non-singlet $'-'$ splitting function,
$a_s \equiv a_s(Q^2)$ and $a_0 = a_s(Q_0^2)$, with
$Q_0^2$ being the input scale. The matrices $\Uvec_1$ and $\Lvec$ are
evolution matrices, for details see Ref.~\cite{BV}. We refrain from applying
so-called small $x$ resummations~\cite{SX,BV1}, since they are very sensitive 
to several series of less singular terms \cite{BV,BV1}, which are yet unknown. Furthermore, no 
factorization theorem exists for these terms
through which non-perturbative and perturbative contributions can be separated in a  well defined way.
Likewise, no other evolution equation than that governing mass singularities
exists to deal with these terms.   

Due to the factor--structure in Eqs.~(\ref{eqNSS}) and (\ref{eqSIS}), the Gaussian 
error propagation of the input density parameters can be calculated analytically, cf.~\cite{BB},
for the whole $Q^2$ region. 
The covariance matrix of the parton distributions alone is completely determined by the fit to the 
data at the input scale.

The input distributions $\Delta_{p,n}^{\rm NS}(N,a_0)$, $\Delta
\Sigma(N,a_0)$ and  $\Delta G(N,a_0)$ are evolved to the
scale $Q^2$. An inverse
{ Mellin}--transform to $x$--space is then performed by a contour
integral in the complex plane around all singularities on the real axis for $x \leq x_0 \leq 1$, which can be
written as
\beq
\label{invers}
\Delta f(x) = \frac{1}{\pi} \int_0^{\infty} dz~{\sf Im} \left[
\exp(i\phi) x^{-c(z)} \Delta f[c(z)]\right].
\eeq
In practice an integral 
along the path $c(z) = c_1 + \rho[\cos(\phi) + i \sin(\phi)]$, with 
$c_1 = 1.1, \rho \ge 0$ and $\phi = (3/4) \pi$ is performed. 
The upper bound on $\rho$ has to be chosen in accordance with the
numerical convergence of the integral~(\ref{invers}) in practice. The result
$\Delta f(x)$ for the respective distribution depends on the
parameters of the spin--dependent parton distributions chosen at the
input scale $Q_0^2$, which are determined by a fit to the data and to
$\Lambda_{\rm QCD}$ or
$\alpha_s(M_Z^2)$, respectively. 
\section{Data Analysis}
\label{sec:data}

\vspace{2mm}
\noindent
The QCD analysis being performed in the following is based on the spin-dependent 
structure functions ${g}_1^{p,d,n}(x,Q^2)$. These structure functions are extracted
from the experimental  cross section asymmetries for longitudinally polarized
leptons scattered off longitudinally polarized nucleons,
\beq
\label{Apara}
A_{||} = \frac 
{\sigma^{{\spar}} - \sigma^{{\sant}}}
{\sigma^{{\spar}} + \sigma^{{\sant}}}~.
\eeq
The arrows $\spar$($\sant$) denote parallel (anti--parallel) relative
spin orientation of the incoming lepton and nucleon.  
The  structure function ratio ${g}_1/F_1$ and the longitudinal
virtual--photon asymmetry $A_1$ are related to $A_{||}$ by 
\beq
\label{asym}
\frac{{g}_1}{F_1} = \frac{1}{(1 + \gamma^2)} \left[ \frac{A_{||}}{D}
+ (\gamma - \eta) A_2 \right]
\eeq
and
\beq
A_1 = \frac{A_{||}}{D} - \eta A_2~,
\eeq
with
\beq
\label{g1F1A12}
\frac{{g}_1}{F_1} 
= \frac{1}{(1 + \gamma^2)} \left[ A_1 + \gamma A_2 \right].
\eeq
The asymmetry $A_2$ is the transverse virtual--photon asymmetry and
constitutes only a small correction to $g_1$. Its contribution has
been treated differently by various experiments as will be
discussed below. The other variables are given by
\bea
\label{D}
D &=& \frac{1-(1-y)\epsilon}{1+\epsilon R(x,Q^2)}~, \\
\label{gamma}
\gamma &=& \frac{2Mx}{\sqrt{Q^2}}~, \\
\label{eta}
\eta &=& \frac{\epsilon \gamma y}{1-\epsilon(1-y)}~, \\
\label{epsilon}
\epsilon &=& \frac{4(1-y)-\gamma^2 y^2}{2y^2+4(1-y)+\gamma^2 y^2}~.
\eea
Here $D$ denotes the virtual photon depolarization 
factor. It determines the fraction of the incoming lepton polarization
transferred to the virtual photon. The variables $\epsilon, \gamma$ and
$\eta$ are kinematic factors, $M$ denotes the mass of the nucleon
and $y=(E-E')/E$ is a Bjorken scaling variable which describes the normalized 
energy transfer to the virtual photon, with $E$ the incoming energy and $E'$ 
the energy of the scattered lepton  in the target rest frame. 
Finally, $R$ denotes the ratio of the longitudinal and transverse
virtual--photon absorption cross section $R(x,Q^2) =
\sigma_L/\sigma_T$, which is experimentally well determined in the kinematic 
region considered in the present analysis. 

In order to obtain ${g}_1(x,Q^2)$ the measured ratio ${g}_1/F_1$
has to be multiplied by the spin--independent structure function
$F_1(x,Q^2)$:  
\beq
\label{g1}
g_1(x,Q^2)=\left(\frac{g_1 }{F_1}\right)(x,Q^2)~\times ~F_1(x,Q^2)~.
\eeq
\noindent
The structure function $F_1(x,Q^2)$ 
can be calculated from the structure function
$F_2(x,Q^2)$ by   
\beq
\label{F1}
F_1(x,Q^2) = \frac {(1+\gamma^2)}{2x(1+R(x,Q^2))}F_2(x,Q^2)~.
\eeq
For $R(x,Q^2)$ and $F_2(x,Q^2)$ parameterizations of existing
measurements are available as will be discussed below.


The following data sets have been used in the present analysis: 
the EMC proton data \cite{EMCp}, the E142 neutron data \cite{E142n},
the {\sc HERMES} neutron data \cite{HERMn}, the E154 neutron data
\cite{E154QCD,E154n}, the SMC proton and deuteron data \cite{SMCpd},
the E143 proton and deuteron data \cite{E143pd}, the {\sc HERMES}
re-analyzed  proton and the new deuteron data \cite{HERMpd}, the E155
deuteron data \cite{E155d}, the E155 proton data \cite{E155p}, the
COMPASS deuteron data \cite{COMPd}, the JLAB neutron
\cite{JLABn}, proton and deuteron data \cite{CLA1pd,CLA2pd}, and the COMPASS proton data 
\cite{COMP1}
\footnote{Earlier data from Ref.~\cite{E80130} are not considered.}.  
The number of data points with  $Q^2 \gsim 1.0~{\rm GeV^2}$
and $W^2 \gsim 3.24~{\rm GeV^2}$ from the different data sets are
summarized in Table~1 together with the $x$ and $Q^2$
ranges of the different experiments. In order to obtain the best possible
statistical accuracy data on $A_1$, $g_1/F_1$ and $g_1$ are not averaged
over the different $Q^2$ values measured within a certain $x$ bin.
In total 1385 data points are used. 
Using $A_1$ and $g_1/F_1$ data has, in addition to the higher
statistics, the advantage of calculating $g_1$ for all these  
data sets in a unique way. Furthermore, $g_1$ data are sometimes
only published as obtained from the average of asymmetries measured at
different $Q^2$ values, while for the QCD analysis it is important to
maintain the $Q^2$ dependence of the measured quantities. 


The SLAC parameterization $R_{1990}$~\cite{R1990} is used 
by most of the experiments when extracting $g_1$. At the time of the
EMC experiment this  
parameterization was not available yet and $R$ was assumed to be $Q^2$ 
independent. SMC adopted a combination of $R_{1990}$ (for $x > 0.12$)
and a parameterization derived by NMC~\cite{RNMC} (for $x < 0.12$). In
the E155 experiment a recent SLAC parameterization for
$R$, $R(1998)$~\cite{RE143}, was used. The changes in the data    
caused by using the different $R$--parameterizations, however,  are not
significant and stay within the experimental errors~\footnote{The EMC
proton data, where the biggest impact is expected, change by a few
percent only, see Ref.~\cite{AAC}.}.
For all $A_1$ and $g_1/F_1$ data sets entering the present QCD
analysis the SLAC $R_{1990}(x,Q^2)$~\cite{R1990} and the NMC
$F_2(x,Q^2)$--parameterization~\cite{F2NMC} is used to perform the
calculation of ${g}_1$. The same parameterizations were used by 
the E154 experiment while JLAB applied the $R(1998)$ SLAC
parameterization of $R$.  


The magnitude of $A_2$ has been measured by SMC~\cite{SMCA2}, 
E154~\cite{EE154}, E143~\cite{E143pd}, E155x~\cite{E155x} and
JLAB~\cite{JLABn} and was found to be small.  
Its contribution to ${g}_1/F_1$ and $A_1$ is further suppressed by
the kinematic factors $\gamma$ and $\eta$ and could in principle be
neglected to a good approximation. On the other hand all the
measurements have shown that the $A_2$ contribution can be
approximated by the Wandzura--Wilczek expression \cite{WW}, which is
calculated from the spin-dependent structure function
${g}_1(x,Q^2)$ assuming that twist--2 contributions are dominant 
according to  
{\footnote{Note that this relation holds also in the presence of
quark and target mass corrections \cite{BT,PR,BLKO}, for non-forward scattering \cite{BR1}, for 
diffractive scattering \cite{BR}, and the gluonic contributions to heavy flavor production 
\cite{BRN}.
Related integral relations for twist--3 contributions  and structure
functions with electro--weak couplings were derived in
Refs.~\cite{BT,BLKO}.} 
\bea
\label{WW}
 A_2(x,Q^2) & = & \frac{\gamma(x,Q^2)}{F_1(x,Q^2)}\Bigl[ g_1(x,Q^2) +
g_2(x,Q^2) \Bigr] \nonumber\\ 
            & \stackrel{WW}{\simeq} & \frac{\gamma(x,Q^2)}{F_1(x,Q^2)}
\int_x^1 \frac{dz}{z} g_1(z,Q^2)~.
\eea
The E143 experiment has exploited its measurement of $A_2$ at 29.1~GeV
and used the Wandzura--Wilczek expression to account for $A_2$
for the other two lower beam energies. The measurement of $A_2$ by
E154 and E155x was done after having published the data on $A_1$ and
was therefore not available for a $A_2$ correction of $A_1$. While
E154 neglected $A_2$, the E155 experiment has used the
Wandzura--Wilczek approximation throughout its data. The JLAB
measurement of $A_2$ went into the extraction of $g_1^n$.  
In {\sc HERMES} the $A_2$ contribution to $g_1^{p,d}/F_1^{p,d}$ data
has been accounted for by using a parameterization for $A_2$ obtained
from a fit $A_2=CM_px/\sqrt{Q^2}$ to all available proton and deuteron
data, 
~\cite{HERMpd}.      
For all $A_1$ data sets used in this analysis $g_1$ has been
calculated with the application of the Wandzura--Wilczek correction
for $A_2$.  

The target mass corrections to the structure function $g_1(x,Q^2)$ are given by \cite{BT,BT1}
\begin{eqnarray}
g_1^{\sf TM}(x,Q^2) &=&
\frac{x}{\xi} \frac{g_1(\xi,Q^2)}{(1+4 M^2 x^2/Q^2)^{3/2}}
+ \frac{4 M^2 x^2}{Q^2} \frac{x+\xi}{\xi (1+4 M^2 x^2/Q^2)^2} \int_\xi^1 \frac{d\xi_1}{\xi_1} 
  g_1(\xi_1,Q^2)
\nonumber
\end{eqnarray}
\begin{eqnarray}
& &
- \frac{4 M^2 x^2}{Q^2} \frac{2 - 4 M^2 x^2/Q^2}{2 (1+4 M^2 x^2/Q^2)^{5/2}} \int_{\xi}^1 
  \frac{d\xi_1}{\xi_1}
  \int_{\xi_1}^1 \frac{d\xi_2}{\xi_2}
  g_1(\xi_2,Q^2)~.
\label{eq:TM}
\end{eqnarray}
Here $M$ denotes the nucleon mass and
\begin{eqnarray}
\xi = \frac{2x}{1 + (1+ 4M^2 x^2/Q^2)^{1/2}}
\end{eqnarray}
is the Nachtmann variable.
We refrain to use approximations to (\ref{eq:TM}) in terms of power series in $M^2/Q^2$ since
the convergence of the corresponding series is problematic as being lined out in Ref.~\cite{BT}. 

The data sets contain both statistical and systematic errors. 
It is known that the systematic errors are partly correlated, which
would lead to an overestimation of the errors when added in quadrature
with the statistical ones, and hence to a reduction of the $\chi^2$
value in the fitting procedure. To treat all data sets on the
same footing only statistical errors were used. 
However, a relative normalization shift, ${\cal{N}}_i$, between the
different data sets was allowed within the normalization
uncertainties, $\Delta {\cal{N}}_i$, quoted by the experiments.  
These normalization shifts were fitted once and then fixed, see
Table~1. 
Thereby the main systematic uncertainties coming from the measurements
of the luminosity and the beam and target polarization were taken into
account.    
The normalization shift for each data set enters as an
additional term in the $\chi^2$--expression for the fit which then
reads 
\bea
\label{CHI2}
\chi^2 \!\!&=&\!\! \sum_{i=1}^{n^{exp}} \left[ 
         \frac {({\cal{N}}_i - 1)^2}
               {(\Delta {\cal{N}}_i)^2} +
           \sum_{h,k=1}^{n^{data}} 
         {({\cal{N}}_i g_{1i,h}^{data} - g_{1,h}^{theor})}
               ({\mathcal{C}}^{-1}_i)^{hk}
         {({\cal{N}}_i g_{1i,k}^{data} - g_{1,k}^{theor})}
         \right],\nn
\eea
where the sums run over all data sets and in each data set over all data 
points. The covariance matrices $\mathcal{C}_i$ are diagonal for each
experiment except for the case of the HERMES data \cite{HERMpd}. The
statistical errors, and consequently the covariance matrices $\mathcal{C}_i$,
have been rescaled by the normalization factors ${\cal{N}}_i$.
The minimization of the $\chi^2$ value above to determine the best 
parameterization of the polarized parton distributions is performed using the 
program {\tt MINUIT} \cite{MINUIT}.
Only fits giving a positive definite
covariance matrix at the end have been accepted in order to be able to 
calculate the fully correlated $1\sigma$ statistical error bands.
\section{Parameterization of the Polarized Parton Distributions}
\label{sec:param}

\vspace{2mm}
\noindent

The shape chosen for the parameterization of the polarized parton
distributions $\Delta f_i(x,Q^2)$ in {$x$--space} at the input scale
$Q^2_0$ is :   
\beq
\label{param}
x\Delta f_i(x,Q_0^2) = \eta_i A_i x^{a_i} (1 - x)^{b_i}
\left(1 + \rho_i x^{\frac{1}{2}} + \gamma_i x\right)\,.
\eeq
The normalization constant $A_i$, being given by 
\bea
\label{anorm}
A_i^{-1} & = & \left( 1 + \gamma_i\frac{a_i}{a_i+b_i+1} \right)
               B(a_i,b_i+1)  
          + \rho_i B\left(a_i+\frac{1}{2},b_i+1\right)~,
\eea
is calculated such that 
\bea
\eta_i = \int_0^1dx \Delta f_i(x,Q_0^2)
\eea
is the first moment of $\Delta f_i(x,Q_0^2)$ at the input scale.  
Here, $B(a,b)$ is the Euler Beta--function being related to
the $\Gamma$--function by $B(a,b)={\Gamma(a)\Gamma(b)}/{\Gamma(a+b)}$.  
The choice of the shape 
(\ref{param}) is applied in various QCD analyses of
unpolarized data, see e.g. Ref.~\cite{CTEQ}.  
The term $x^{a_i}$ controls the behavior of the parton density at
low and $(1 - x)^{b_i}$ that at large values of $x$, respectively. The
remaining polynomial factor accounts for potential degrees of freedom
at medium $x$. At the same time not too many
parameters can be sufficiently constrained by the data available at present. In particular 
the parameters $\rho_i$, which play a role in unpolarized analyses, \cite{BBG}, are set to zero, 
cf. also \cite{BB,DSSV}.

If the QCD evolution equations are solved in {Mellin} space as
described in Section~\ref{sec:qcdana} a {Mellin} transformation of
the polarized parton density $\Delta f(x,Q^2)$ is performed for 
complex arguments $N$~:
\bea
\Mvec[ \Delta f_i(x,Q_0^2)](N)  &=& \int_0^1 x^{N-1} dx \Delta
f_i(x,Q^2_0)  \nonumber\\ 
   &=& \eta_i A_i 
    \left( 1 + \gamma_i\frac{N-1+a_i}{N+a_i+b_i} \right)
                   B(N-1+a_i,b_i+1)  
~.
\label{MNmom}
\eea
In the present analysis we assume (approximate) SU(3) flavor symmetry the sea quark distribution is 
given by
\bea
\label{eq:sq}
\Delta \overline{q}_s(x,Q^2) &=& \Delta \overline{u}(x,Q^2) = \Delta
\overline{d}(x,Q^2) =
\Delta {s}(x,Q^2) = \Delta \overline{s}(x,Q^2)~, \\
\Delta \Sigma(x,Q^2) &=& \Delta u_v(x,Q^2) + \Delta d_v(x,Q^2) + 6 \Delta \overline{q}_s(x,Q^2)~.
\eea
\noindent
We refer to the inclusive polarized DIS World Data only.
A breaking of the flavor symmetry also for the light (sea) quarks in the 
polarized case
is probable and has been clearly observed in the unpolarized case, cf. e.g. \cite{ABKM}.
In the polarized case first attempts have been made to determine the individual 
sea quark distributions, cf.~Refs.~\cite{HERMES1,DSSV}. We consider the evolution 
of the complete light polarized sea.

Under the above assumption four spin--dependent parton densities have to be determined in the QCD
analysis. They are chosen to be:  
$\Delta u_v(x,Q^2)$, $\Delta d_v(x,Q^2)$, $\Delta
\overline{q}_s(x,Q^2)$ and $\Delta G(x,Q^2)$.
As seen from Eq.~(\ref{param}), each spin--dependent density contains
five parameters which gives a total of 20 for all four. It has been
found that, in order to meet the quality of the available data and the
reliability of the fitting procedure, this large number of free
parameters has to be reduced, which is discussed below.

The first moments of the polarized valence distributions $\Delta u_v$
and $\Delta d_v$, $\eta_{u_v}$ and $\eta_{d_v}$, can be fixed
exploiting the knowledge of the  parameters $F$ and $D$ as 
measured in neutron and hyperon $\beta$--decays according to the
relations: 
\bea
\eta_{u_v} - \eta_{d_v} & = & F + D~, \\
\eta_{u_v} + \eta_{d_v} & = & 3F - D~.
\eea
A re--evaluation of $F$ and $D$ was performed on the basis of updated
$\beta$--decay constants \cite{PDG} leading to 
\beq
\label{fd}
F=0.464 \pm 0.008 \quad {\rm and} \quad D=0.806 \pm 0.008~,
\eeq
and consequently to
\beq
\label{etauvdv}
\eta_{u_v} = 0.928 \pm 0.014 \quad {\rm and} \quad \eta_{d_v} =
-0.342 \pm 0.018~. 
\eeq
In order to compensate for the present insufficient accuracy of the
data, a certain number of parameters is set to zero after this was thoroughly suggested by initial 
fits. This applies to
$\rho_{u_v} = \rho_{d_v} = 0$, $\gamma_{\overline q_s} =  
\rho_{\overline q_s} = 0$, and $\gamma_G = \rho_G = 0$. 
The number of parameters to be fitted for each
polarized parton density is reduced to three, i.e. to 12 in total.
In addition the QCD scale $\Lambda_{\rm QCD}$ is fitted.

In the analysis it turns out that the
four parameters $\gamma_{u_v}$, $\gamma_{d_v}$, $b_{\overline q_s}$,
and $b_G$ have very large uncertainties. The precision of the data 
is not high enough to constrain these parameters sufficiently.
Altering them within these uncertainties does not lead to a
significant change of $\chi^2$. These four parameters were therefore
fixed. The first two of them were fixed at their values obtained in the
initial fitting pass,  $\gamma_{u_v} = 27.64$ and
$\gamma_{d_v} = 44.26$.
In fixing the high--$x$ slopes $b_G$ and $b_{\overline q_s}$  a
relation was adopted as derived from the unpolarized parton densities,
$b_{\overline q_s}/b_G{\rm (pol)} = b_{\overline q_s}/b_G {\rm
(unpol)} = 1.44$. Fitting with this constraint led to the following
choice: $b_G = 5.61$ and $b_{\overline q_s} = 8.08$, see
e.g. Ref.~\cite{GRVr}.

A second relation was adopted to constrain the low--$x$ behavior of the
spin--dependent gluon density with respect to the low--$x$ behavior of
the spin--dependent sea--quark distribution by 
$a_G = a_{\overline q_s} + C$ with $C = 1$.
This relation, together with the relation for the high--$x$ slopes, are
suited to establish positivity for $\Delta G$ and $\Delta {\overline
q_s}$. No explicit positivity constraint has been enforced for
$\Delta u_v$ and $\Delta d_v$.
\section{Determination of the Errors}
\label{sec:error}

\vspace{2mm}
\noindent
\subsection{Calculation of Statistical Errors}
\label{sec:staerr}

The evolved polarized parton densities and structure functions are
functions of the input densities. Let $\Delta f(x,Q^2;p_i|_{i=1,k})$
be the evolved polarized density at the scale $Q^2$ depending on the parameters
$p_i|_{i=1,k}$. Then its correlated statistical error as given by
Gaussian error propagation is 
\beq
\label{errprop}
(\sigma \Delta f)^2 =  \sum_{i,j=1}^k 
                \left( \frac{\partial \Delta f}{\partial p_i}
                \frac{\partial \Delta f}{\partial p_j} \right)
                \tx{cov}(p_i,p_j)\,, 
\eeq
where $\tx{cov}(p_i,p_j)$ are the elements of the covariance matrix
determined in the QCD analysis. The gradients $\partial \Delta
f/\partial p_i$ at the input scale $Q_0^2$ can be calculated analytically. 
Their values at $Q^2$ are evaluated through the  evolution and can then be used
to compute  the errors according to Eq.~(\ref{errprop}). As shown in
Section~\ref{sec:qcdana}, the covariance matrix is completely determined
by the fit at the input scale and does not change when the evolution
is done in {Mellin} space. That means it can be used at any scale
$Q^2$. For the expressions of the gradients at the input scale see 
Ref.~\cite{BB}.

Apart from statistical errors the data are also subject to systematic
uncertainties which are even partly correlated. These correlations are
not published by the experiments and can hence not be taken into
account here. In the following the influence of experimental and
theoretical systematic uncertainties will be investigated.

\subsection{Determination of Experimental Systematic Uncertainties}
\label{subsec:sysexp}

The experimental systematic uncertainties were estimated from the
following sources~:   

\begin{itemize}

\item[1.] {\sf The variation of the data within their experimental
systematic uncertainties}.  \\
The procedure used to obtain
the contribution from the experimental systematic uncertainties
consists in shifting each data set by 
$\pm\sigma_{syst}$ while leaving the other data sets at their central
values and looking at how much the polarized distributions change. The
extreme changes from the `central' distribution were taken as the
systematic uncertainties. This was done for each of the 19 data sets
used separately and, finally, the 19 contributions were added in
quadrature to obtain the total contribution. 

\item[2.] {\sf  The variation of the data within the upper and
lower limits of the NMC $F_2$ parameterization}~\cite{F2NMC}. \\  
When calculating ${g}_1(x,Q^2)$ from the asymmetry data, see Section~\ref{sec:data}, 
the NMC $F_2$--parameterization was used
at its upper and lower limit to determine the changes in the polarized
distributions compared to the `central' curve. The extreme deviations
from that curve were taken as the systematic uncertainties arising
from the $F_2$ parameterization. 

\item[3.] {\sf  The variation of the data within the uncertainty of
the $R$ parameterization }~\cite{R1990}. \\
When calculating
${g}_1(x,Q^2)$ from the asymmetry data, see Section~\ref{sec:data}, the
SLAC $R_{1990}$ parameterization was used 
at its uncertainty limits, see Ref.~\cite{R1990}, to determine the
changes in the spin--dependent distributions compared to the `central'
curve. The maximal deviations from that curve were taken as the
systematic uncertainties arising from the $R$--parameterization.

\end{itemize}

\noindent
Finally, the contributions from all three sources were added in
quadrature to obtain the total contribution at each value of $x$, which is shown
as hatched error bands in the Figures below.

\subsection{Determination of Theoretical Systematic Uncertainties}
\label{subsec:systhe}

The theoretical systematic uncertainties were estimated from the
following sources~:   

\begin{itemize}

\item[1.] {The variation of the factorization and the
renormalization scale by a factor of 2}.  

\item[2.] {An additional variation of $\Lambda_{QCD}^{(4)}$.} We varied 
$\Lambda_{QCD}^{(4)}$ by $\pm 30$ MeV, which corresponds to a { variation of
$\alpha_s(M_Z^2)$} by $\pm 0.002$, being a typical error in individual precision measurements,
in addition to the error of $\alpha_s(M_Z^2)$ being determined in the present 
fit.

\item[3.] { The variation of $\eta_{u_v}$ and $\eta_{d_v}$}
within the errors of the parameters $F$ and $D$, see
Section~4,  while keeping $g_A/g_V = F + D$ constant.

\item[4.] {\sf The variation of the parameterizations at the input
scale $Q_0^2$}.\\ 
Two cases are considered: first, the values of the
parameter $\rho = -1.0$ for $\Delta u_v$ and $\Delta d_v$ are added in the
fit. Second, the values of the
parameter $\gamma = 17.64$  for $\Delta u_v$ and 24.26 for $\Delta d_v$, respectively, for the same 
densities are changed compared to the
values used for the central curve.  

\item[5.] {The variation of the standard input scale:} from
$Q_0^2 = 4.0 ~{\rm GeV^2}$ to $2.0~{\rm GeV^2}$ and $6.0~{\rm GeV^2}$. 

\end{itemize}

For all items always the extreme deviations from the {`central'}
curve were taken as the contribution to the theoretical systematic
uncertainty from that source.   
Finally, all different contributions were added in quadrature to get
the total contribution at each value of $x$ which is shown as the second 
hatched error bands in the Figures below. 

\section{Results of the QCD Analysis}
\label{sec:results}
\subsection{The parton distributions}
\label{sec:pdfs}

\vspace{2mm}
\noindent
In the NLO QCD fit we covered the polarized world deep-inelastic data 
with $Q^2 > 1~\GeV^2, W^2 > 3.24~\GeV^2$, with a value of $\chi^2/NDF = 1.12$.
We have chosen a value {\tt UP = 9.3} corresponding to the 7+1 parameter fit in the {\tt 
MINUIT}-procedure.
The parton distributions are parameterized at $Q_0^2 = 4~\GeV^2$. In Table~2 the 
values of the fit parameters are summarized. The covariance matrix
of the fit is given in Table~3. Gaussian error propagation allows to derive 
the error bands due to the parton densities      
for the various polarized observables. 

In Figure~1 the polarized momentum distributions $x \Delta f_i(x)$ are presented
at the scale $Q_0^2$. We compare with other analyses \cite{AAC,GRSV,LSS,DSSV}.
The $x \Delta u_v$ distribution is slightly lowered if compared to our previous analysis 
\cite{BB}. The most important change concerns the gluon distribution  $x \Delta G$, 
which is lowered by about a factor of two relative to the results of Ref.~\cite{BB}. Comparing to 
the results of other analyses $x \Delta u_v$ turns out to be lower in a wider range. 
In case of the 
$x \Delta d_v$-distribution all fits widely agree within the 1$\sigma$ error band.
This also applies for the $x \Delta \overline{q}$-distribution. 
For the polarized gluon distribution the agreement of 
the different fits at larger values of $x$ agree within the 1$\sigma$ error, while
below $x \sim 0.02$ the fits \cite{GRSV,AAC,LSS} yield slightly higher values.
We indicated the positivity constraints referring to the unpolarized PDFs \cite{MSTW}, which are 
obeyed.

In Figures~2 and 3 we analyze the systematic errors in more detail, cf.~Sections~5.2
and 5.3. For the gluon distribution function $x \Delta G(x,Q_0^2)$, Figure~2, the experimental 
systematic errors are lower than the statistical errors, but are still significant.
The combined theoretical systematic effects at NLO amount to larger values than
the experimental ones. About half of the error is due to the uncertainty in
$\Lambda_{\rm QCD}$. Clearly, in future analyses based on NNLO QCD evolution
this error will diminish. The effect of the experimental and theoretical systematic errors
in case of the singlet distribution $x \Delta \Sigma(x,Q^2_0)$ is similar. The errors
are smaller if compared to the gluon distribution. The singlet distribution
at $Q_0^2 = 4~\GeV^2$ is negative (within the 1$\sigma$ errors) for $x < 2 \cdot 10^{-2}$ and turns
to positive values for $x > 4 \cdot 10^{-2}$. All the fits \cite{GRSV,AAC,LSS} lie inside the
1$\sigma$ error band. 
In the medium $x$-range the DSSV distribution \cite{DSSV} 
yields somewhat larger values.

In Figure~4 we compare the fit results for the structure functions 
$g_1^p(x,Q^2), g_1^d(x,Q^2)$ and $g_1^n(x,Q^2)$ at $Q^2 = 5~\GeV^2$ with the data, cf.~Table~1, 
to illustrate the fit quality for the different targets as an example. Furthermore, also the 
results of the GRSV \cite{GRSV} and AAC \cite{AAC} analyses are shown. 
Overall a good agreement is obtained.

A further illustration of the fit quality is presented in Figure~5. Here the data for $g_1^p(x,Q^2)$
are compared to the fit including, the statistical errors. We also show the fit results of 
AAC, GRSV, and LSS~\cite{AAC,GRSV,LSS}. Within the error bands the data agree well with the fit.
In the lowest $x$-bins the fluctuation is somewhat larger. In some  cases the EMC-data  \cite{EMCp} lay
outside the 1$\sigma$ error range. 
 
\subsection{\boldmath $\Lambda_{\rm QCD}$ and $\alpha_s(M_Z^2)$}
\label{sec:lambda}

\vspace{2mm}
\noindent
The NLO QCD-analysis of the polarized world deep-inelastic data requires
the fit of $\Lambda_{\rm QCD}$ along with the parameters of the non-perturbative
parton distributions at the scale $Q_0^2$. As outlined in Table~3 the QCD-scale is
correlated to all other parameters and in particular also to $\eta_G$. Due to this
analyses in which $\alpha_s(M_Z^2)$ or $\Lambda_{\rm QCD}$ is imported from 3rd
sources may suffer significant biases, i.e. a too {\it large} value of $\alpha_s(M_Z^2)
(\Lambda_{\rm QCD})$ leads to a too {\it small} gluon distribution, aside of other
effects. Despite of various precision measurements of the strong coupling constant
based on theoretical NNLO (and partially even higher) precision, a thorough agreement
on the value of $\alpha_s(M_Z^2)$ has not yet been reached, cf. e.g.~\cite{BETHKE}.
Due to this $\Lambda_{\rm QCD}$ is determined in this analysis. We refer to
$\Lambda_{\rm QCD}^{{\rm NLO}, (N_f = 4)}$ as the NLO value for 4 active flavors and
obtain 
\begin{eqnarray}
\Lambda_{\rm QCD}^{(4)} = 243 \pm 62 (\rm exp)~~\MeV~.
\end{eqnarray}
In an earlier analysis \cite{BB} the values
\begin{eqnarray}
\Lambda_{\rm QCD}^{(4)} = 235 \pm 53 (\rm exp)~~\MeV~{\tt ISET = 3} \\
\Lambda_{\rm QCD}^{(4)} = 240 \pm 60 (\rm exp)~~\MeV~{\tt ISET = 4} 
\end{eqnarray}
were found, for comparison, slightly depending on some assumptions in the fit.
The variation of the factorization and renormalization scales $\mu_{f,r}^2$ 
by a factor of 1/2 and 2, respectively, yields
\begin{eqnarray}
\Lambda_{\rm QCD}^{(4)} = 243 \pm 62~~(\rm exp)~~\begin{array}{l} -37 \\ +21 \end{array}~{\rm (FS)}  
~~\begin{array}{l} +46 \\ -87 \end{array}~{\rm (RS)}  
~~\MeV~.
\end{eqnarray}
Here we excluded values $\mu_{f,r}^2 < 1~\GeV^2$, unlike in Ref.~\cite{BB}, since at
scales lower than 1~$\GeV^2$ the perturbative description cannot be considered reliable
anymore.

Correspondingly, for $\alpha_s(M_Z^2)$ one obtains
\begin{eqnarray}
\alpha_s(M_Z^2)  = 0.1132 
~~\begin{array}{l}  + 0.0043 \\ -0.0051 \end{array}~{\rm (exp)}  
~~\begin{array}{l}  - 0.0029 \\ +0.0015 \end{array}~{\rm (FS)}  
~~\begin{array}{l}  + 0.0032 \\ -0.0075 \end{array}~{\rm (RS)}~,  
\end{eqnarray}
with combined errors of
\begin{eqnarray}
\label{curalp}
\alpha_s(M_Z^2)  = 0.1132 
~~\begin{array}{l}  + 0.0056 \\ -0.0095 \end{array}~.  
\end{eqnarray}
Due to the NLO analysis the factorization- and renormalization scale uncertainties
are still dominant. The values are well compatible with recent determinations 
of the strong coupling constant at NNLO and N$^3$LO from deep-inelastic data~:
\begin{eqnarray}
\alpha_s(M_Z^2)  &=& 0.1134 
                           {\footnotesize \begin{array}{l} + 0.0019 \\
                                            - 0.0021 \end{array}} \hspace*{6mm} {\rm NNLO}  
\hspace*{36.5mm} [28] \\
\alpha_s(M_Z^2)  &=& 0.1141 {\footnotesize
                           \begin{array}{l} + 0.0020 \\
                                            - 0.0022 \end{array}} \hspace*{6mm} {\rm N^3LO}    
\hspace*{38mm}
[28] \\
\alpha_s(M_Z^2)  &=& 0.1135 \pm 0.0014 \hspace*{5mm} {\rm NNLO,~FFS}                            
\hspace*{26mm}
[31] \\
\alpha_s(M_Z^2)  &=& 0.1129 \pm 0.0014 \hspace*{5mm} {\rm NNLO,~BSMN}                           
\hspace*{21.5mm}
[31] \\
\alpha_s(M_Z^2)  &=& 0.1124 \pm 0.0020 \hspace*{5mm} {\rm NNLO,~dyn.~approach}                  
\hspace*{8mm}
[29] \\
\alpha_s(M_Z^2)  &=& 0.1158 \pm 0.0035 \hspace*{5mm} {\rm NNLO,~stand.~approach}                
\hspace*{5mm}
[29] \\
\alpha_s(M_Z^2)  &=& 0.1171 \pm 0.0014 \hspace*{5mm} {\rm NNLO}                                 
\hspace*{37mm}
[30] 
\end{eqnarray}
More recent unpolarized NNLO fits, including the combined HERA data~\cite{COMH1Z}, yield
\begin{eqnarray}
\alpha_s(M_Z^2)  &=& 0.1147 \pm 0.0012 \hspace*{5mm} {\rm NNLO}  \hspace*{2.85cm}~[67]
\\
\alpha_s(M_Z^2)  &=& 0.1145 \pm 0.0042 \hspace*{5mm} {\rm NNLO,~ (preliminary)}~~[68]
\end{eqnarray}
The central value of the present fit (\ref{curalp}) does well compare with the above values.
They are located below the present weighted average of $\alpha_s(M_Z^2)$ measurements \cite{BETHKE} of
\begin{eqnarray}
\label{alpWA}
\alpha_s(M_Z^2)  = 0.1184  \pm 0.0007~.
\end{eqnarray}
The error given in (\ref{alpWA}) does not include any yet unknown relative systematics between the 
different classes of the same type of measurement. 

We would like to mention that recent determinations of $\alpha_s(M_Z^2)$ using event shape
moments for high energy $e^+e^-$ annihilation data from PETRA and LEP including power corrections
the following values were obtained~:
\begin{eqnarray}
\alpha_s(M_Z^2)  &=& 0.1135 \pm 0.0002~{\rm (exp)} \pm 0.005~{(\Omega_1)} \pm 0.0009~{\rm (pert)} 
\hspace*{5mm} {\rm NNLO}
 \hspace*{5mm}     
[69]\\
\alpha_s(M_Z^2)  &=& 0.1153 \pm 0.0017~{\rm (exp)} \pm 0.0023~{\rm (th)} \hspace*{32.5mm} {\rm NNLO}    
 \hspace*{5mm} [70]
\end{eqnarray}
Also these measurements of $\alpha_s(M_Z^2)$ yield low values. They show that the results
obtained analyzing deep--inelastic data do not form a special case. The systematics of the 
different extractions of $\alpha_s(M_Z^2)$ has to be understood in more detail in the future.

Fit results from previous polarized analyses like \cite{SMCQCD,E154QCD,ABFR} were 
summarized in \cite{BB}.
In Figure~6 we compare recent determinations at NNLO and N$^3$LO for unpolarized 
and at NLO for polarized deep-inelastic scattering. 
\subsection{Moments of Polarized Parton Distributions}
\label{sec:moments}

\vspace{2mm}
\noindent
We calculate the lowest moments of the polarized parton densities
\begin{eqnarray}
\langle f(x) \rangle_n = \int_0^1 dx x^n \Delta f(x)~,
\end{eqnarray}
where $\Delta f(x)$ denote the different polarized (number) density 
distributions. The moments $n = 0, ...,3$ are given in Table~4.
The behaviour of these distributions outside the kinematic 
range in which the fit is performed bear uncertainties, which are difficult 
to predict for these non-perturbative quantities.~\footnote{We remind the
failure in predicting the lower  $x$ behaviour of the structure function 
$F_2(x,Q^2)$ prior the HERA measurements until 1992, which assumed a slightly 
falling or constant behaviour below $x \simeq 10^{-2}$, whereas a strong rise
was measured at HERA.} Instead of presenting necessarily uncertain models
for this range, we compute the respective part of the moments for values
$x < 0.005$ and $x > 0.75$ extrapolating the present distributions to the
range $x \in [0,1]$. 

The zeroth moments of the polarized quark- and gluon distributions as well as 
the contributions due to the quark- and gluon angular momenta, $L_q$ and $L_g$,
constitute the nucleon spin
\begin{eqnarray}
\label{eq:spin}
\frac{1}{2} &=& \frac{1}{2} \langle \Delta \Sigma(x) \rangle_0 + \langle \Delta G(x) \rangle_0
+ L_q + L_g~.
\end{eqnarray}
We obtain
\begin{eqnarray}
\langle \Delta \Sigma(x,Q_0^2) \rangle_0 &=& 0.193 \pm 0.075 \\
\langle \Delta G(x,Q_0^2) \rangle_0      &=& 0.462 \pm 0.430~,
\end{eqnarray}
at $Q_0^2 = 4 \GeV^2$. For Eq.~(\ref{eq:spin}) this yields
\begin{eqnarray}
\frac{1}{2} = \left(0.555 \pm 0.436 \right) + L_q + L_g~.
\end{eqnarray}
The error of $\Delta G$ is clearly dominant.

The results given in Table~4 can be compared to ab initio
calculations of these moments in Lattice Gauge Theory. 
There we also compare with the values obtained in our previous analysis~\cite{BB}.
In the present analysis the first moments ($n=0$) of the polarized valence quark 
distributions are determined by the values of $F$ and $D$ and are fixed in the fit.
Comparing to the results of \cite{BB}, {\tt ISET = 3}~\footnote{The NLO results for 
{\tt ISET = 4} are quite similar.} slightly larger values are obtained for the moments
of $\Delta u_v(x,Q^2)$ and slightly lower values for  $\Delta d_v(x,Q^2)$ and 
$\Delta \overline{q}(x,Q^2)$. A very significant change is obtained for the moments of the 
polarized gluon density, where the moments reduced by a factor of about two comparing to
\cite{BB}. Although being positive, the latter moments are now compatible with zero 
in the 1$\sigma$ errors.

First lattice results for the moments $n = 0,1,2$ of the polarized quark distributions 
were given about ten years ago. Meanwhile many systematic effects in the simulation
have been improved further. Still there are differences in the different simulations.
Rather aiming on a detailed comparison with the values in Table~4 we give a brief summary of the 
current status. A recent survey has been presented in \cite{DR}. Lattice results on 
$\langle \Delta u - \Delta d \rangle_0$ for $m_{\pi}^2 = 0.029 ... 0.48\GeV^2$ were given in
\cite{RBC,LAT1,LHPC,QCDSF1} by the BGR, RBC, LHPC, ETMC, QCDSF-collaborations using dynamical quarks. 
Most of the values are yet below the experimental value. The QCDSF collaboration \cite{QCDSF1}
performed simulations at $m_{\pi} = 170 \MeV$ and obtained
\begin{eqnarray}
\langle \Delta u(x) - \Delta d(x) \rangle_0         &=& 1.17 \pm 0.05~, \\
\frac{\langle \Delta u(x) - \Delta d(x) \rangle_0}  
     {\langle \Delta u(x) + \Delta d(x) \rangle_0}  &=& 0.47 \pm 0.02~. 
\end{eqnarray}

For the first moment the following values were determined
\begin{eqnarray} 
\langle \Delta u - \Delta d \rangle_1 
&=  0.271 \pm 0.040, \hspace*{7mm} m_\pi = 493 \MeV,& \hspace*{7mm} [72] \\  
                                      &=  0.252 \pm 0.020, \hspace*{7mm} m_\pi = 352 \MeV,& 
\hspace*{7mm} [74]
\end{eqnarray}
which are larger than the value
\begin{eqnarray} 
\langle \Delta u - \Delta d \rangle_1 &=& 0.190 \pm 0.008  
\end{eqnarray}
determined in the present analysis. 

Results on the second moment were given in \cite{LHPC}
\begin{eqnarray} 
\langle \Delta u - \Delta d \rangle_2 &=&  0.083 \pm 0.012, \hspace*{7mm} m_\pi = 352 \MeV,\\  
\end{eqnarray}
to be compared to
\begin{eqnarray} 
\langle \Delta u - \Delta d \rangle_2 &=& 0.063 \pm 0.004  
\end{eqnarray}
obtained in this analysis. 

Results of older lattice simulations \cite{LAT2}  were discussed 
in \cite{BB} previously. The values given above for the 1st and 2nd moment are based on theoretically
much improved simulations, if compared to early investigations \cite{LAT2}. Yet, the moments  
obtained yield similar values. Comparing the lattice results with the results obtained in QCD-fits
to the polarized deep-inelastic world data one observes a similar trend of values but not yet agreement.

\section{Higher Twist}
\label{sec:htwist}

\vspace{2mm}
\noindent
So far we have applied the twist-2  approximation at NLO to describe 
the spin--dependent structure function ${g}_1$. As the data may contain 
contributions from higher twist (HT) it is of interest to look for possible 
effects of such contributions. A thorough description of higher twist anomalous 
dimensions and Wilson coefficients to NLO is still missing, even for the 
twist--4 contributions. Therefore we will perform a purely phenomenological 
analysis. Similar to the approach of  Ref.~\cite{VirMil} for the structure 
function $F_2$, where a higher twist term is parameterized by the ansatz 
\beq
h_{HT}^i(x,Q^2) = \frac{C_i(x)}{Q^2}, 
\eeq
and used multiplicative to  the leading twist (LT) contribution,  $g_1(x,Q^2)$ 
is described by
\beq
g_{1}^i(x,Q^2) = g_{1, {LT}}^i(x,Q^2)\left[1 + h_{HT,m}^i(x,Q^2)\right]~,
\eeq
where $i = p,d,n$. Likewise onw may describe an additive higher twist term
\beq
g_{1}^i(x,Q^2) = g_{1, {LT}}^i(x,Q^2) +  h_{HT,a}^i(x,Q^2)~.
\eeq
This approach has to be handled with great care, since the coefficients
$C_i(x)$  are actually also $Q^2$ dependent. They consist of a combination of 
various terms which exhibit different scaling violations. The relation to 
$\Lambda_{\rm QCD}$ is completely masked here. Moreover, higher twist 
contributions should have a flavor-dependence and are not expected to be 
the same in case of polarized and unpolarized deep-inelastic scattering.

The kinematic $x$--range being covered by experiment is divided into 5 bins 
and the coefficient $C_i(x)$ has to be determined in each bin and for each
target. The resulting coefficients for the proton and the deuteron target, 
$C_p(x)$ and $C_d(x)$, are summarized in Table~5 both for the multiplicative and additive cases. 
Due to the enlarged number
of parameters being fitted the value of {\tt UP = 11.5} is used.
In the additive and multiplicative cases we observe  quite comparable patterns.
We prefer the additive case, since the twist--2 scaling violations of $g_1(x,Q^2)$ do not 
influence $C_{p,d,n}(x)$.
Here, in most of the bins the result is $1~\sigma$ compatible with zero, except of one case for 
$C_p(x)$. 
$C_{d,n}(x)$  are all $1~\sigma$ compatible with zero. The behaviour
in case of the deuteron and neutron is rather flat, while for the proton a slight structure,
yet with large errors, is indicated, cf. Figure~\ref{fig:ht}. In the fit determining also the higher 
twist parameters the value $\chi^2/NDF$ for the CLAS data amounts to 1.12. 
The present data are not yet precise enough to undoubtedly reveal 
higher twist contributions in the range $Q^2 > 1~{\rm GeV^2}$ and a
NLO QCD analysis can be carried out in the leading twist
approximation. We do not confirm the results of Ref.~\cite{LSS_HT}. 
Unlike the case for the large $x$ valence quark region, in which dynamical 
higher twist terms are extracted consistently in the unpolarized case, 
cf. Refs.~\cite{SA1,BBG,BB1}, the situation is more involved for the lower $x$-region. 
The dynamics is clearly different in both these domains due to the contributing 
parton species. As has been shown in \cite{SA2}, different power 
corrections cancel each other in the small $x$ region.
\section{Conclusions}
\label{sec:concl}

\vspace{2mm}
\noindent
A QCD analysis of the polarized deep-inelastic world data has been performed at NLO,
including the effects of charm production to first order. We derived a parameterization
for the polarized parton distributions and $\Lambda_{\rm QCD}^{N_f =4}$ with the error 
correlations between the fitted parameters applying the $\chi^2$--method. Detailed 
comparisons have been performed with recent parameterizations \cite{AAC,GRSV,LSS,DSSV}.
The present data are not accurate enough to determine all the shape parameters at a
sufficient accuracy. Due to this some of the parameters have to be fixed after 
an initial phase of the analysis to a model-value. If compared to our previous analysis
\cite{BB} the more recent data lead to a smaller gluon distribution, which is for a wide region
of $x$ compatible with zero within the $1\sigma$ error. We determined both the 
experimental and theoretical systematic effects. Both the central values of the parton densities
and their $1\sigma$ error are made available in form of a numerical parameterization in the
range $x \in [10^{-9}, 1],~Q^2 \in [1, 10^6]~\GeV^2$. These distributions can be used
for polarized hard--scattering processes at hadron-- and lepton--nucleon colliders for various observables,
including error propagation w.r.t. the accuracy of the parton densities. The implementation in  terms 
of grid-interpolation is well suited also for Monte Carlo simulations. 

The QCD-scale was determined by $\Lambda_{\rm QCD}^{N_f = 4} = 243 \pm { 62}~~{\rm 
(exp)}~\footnotesize{\begin{array}{l} 
+ {\footnotesize 59}\\ {\footnotesize -90} 
\end{array}}~~(\rm th)~~\normalsize \MeV$,
corresponding to $\alpha_s^{\rm NLO}(M_Z^2)
= 0.1132~~{\footnotesize \begin{array}{l}  {\footnotesize + 0.0056} \\ {\footnotesize -0.0095} \end{array}}$.
The central value is well compatible with other measurements, 
cf.~\cite{BBG,JR,ABKM,ABM,H1ZEUS,HOANG,Gehrmann:2009eh}. The errors are still rather large, also because of 
the scale variation uncertainties at NLO. Nonetheless the correlated determination of $\alpha_s(M_Z^2)$ with
the parton densities is of importance to avoid biases in particular w.r.t. to the size of the gluon distribution
function.

We also determined potential higher twist contributions, which were found to be compatible with zero
in the whole kinematic range within the present errors for $C_{n,d}(x)$ and also for $C_{p}(x)$, 
except 
for one of the bins. 
Based on the results of the present analysis we computed the lowest moments of the individual twist--2
parton densities. For the lowest moment $(1/2)\langle \Delta \Sigma(x) \rangle_0 + \langle \Delta G(x) \rangle_0$
we obtain $0.555 \pm 0.436$ at $Q_0^2~=~4~\GeV^2$, which is well compatible with the nucleon spin 
$1/2$ even
without angular momentum contributions. However, the error is dominated by that of the polarized gluon
distribution.
The moments derived may be compared to upcoming lattice simulations based on dynamical quarks of the 
corresponding operator matrix elements. The present results are not yet in agreement, although the tendency
of values is visible. Runs at even smaller values of $m_{\pi}$ seem to be necessary.  

\vspace{5mm}\noindent
{\bf Acknowledgment.}~
We would like to thank S. Alekhin, E. Aschenauer, L. De Nardo, M.~G\"ockeler, K. Jansen, W.D. Nowak, D. 
Renner, and 
G. Schierholz for discussions.
This work was supported in part by DFG Sonderforschungsbereich Transregio 9, Computergest\"utzte
Theoretische Teilchenphysik and the European Commission MRTN HEPTOOLS under Contract No.
MRTN-CT-2006-035505.

\vspace{2mm}
\noindent

\newpage
\section{Appendix: The {\tt FORTRAN}-code for the parton densities and
their errors}

\vspace{2mm}
\noindent
A fast {\tt FORTRAN} program is available to represent the polarized
parton densities $x\Delta u_v(x,Q^2)$, $x\Delta d_v(x,Q^2)$, $x\Delta
G(x,Q^2)$, and $x\Delta \bar{q}(x,Q^2)$, as well as the polarized structure
functions $xg_1^p(x,Q^2)$ and $xg_1^n(x,Q^2)$ at NLO in the $\overline{\rm MS}$--scheme 
together with the parameterizations of their $1\sigma$  errors.      
The following ranges in $x$ and $Q^2$ are covered:

\vspace*{-0.15cm}

\begin{center}
$10^{-9} < x < 1 \quad , \quad 1~\GeV^2 < Q^2 < 10^6~\GeV^2.$
\end{center}

\vspace*{-0.15cm}

\noindent
The polarized distributions are the result of a fit to the world data
on spin asymmetries, i.e. $A_1^{p,n,d}$ or $g_1/F_1^{p,n,d}$, as
described above.
The {\tt SUBROUTINE POLPDF} returns the values of the polarized
distributions, always multiplied by $x$, at a given point in
$x$ and $Q^2$ by interpolating the data on specified grids. The
interpolation in $x$ is done by cubic splines and in $Q^2$ by a
linear interpolation in $\log\,(Q^2)$.~\footnote{We thank S.~Kumano and
M.~Miyama of the AAC--collaboration for allowing us to use their
interpolation routines.}

The parton distributions are evaluated by
\begin{center}
{\tt
SUBROUTINE POLPDF(ISET, X, Q2, UV, DUV, DV, DDV, GL, DGL, SEA, DSEA, G1P,%
DG1P,G1N,DG1N)},
\end{center}
with {\tt ISET = 1}.
All non-integer variables are of the type {\tt REAL*8}. The calling
routine has to contain the {\tt COMMON/INTINI/ IINI}. Before the first
call to {\tt SUBROUTINE POLPDF} the initialization is set by {\tt IINI = 0}.

The parameters {\tt X, Q$^2$ [\GeV$^2$]} are $x$ and $Q^2$. The momentum
densities of the polarized up- and down valence quarks, gluons and
the sea quarks are {\tt UV, DV, GL, SEA}, with ${\tt SEA}~=~x\Delta
u_s = x\Delta d_s = x\Delta \overline{u} = x\Delta \overline{d}
= x\Delta s = x\Delta \overline{s}$. Correspondingly, {\tt DUV}
is the $1\sigma$  error of {\tt UV} etc. and {\tt G1P} and {\tt G1N}
are the values of the electromagnetic structure functions $xg_1^p$
and $xg_1^n$.

The programme {\tt example.f} reads the data-grid                                                   
{\tt qcd\_nlo\_905\_0.grid} and
is compiled using {\tt gfortran} at a {\tt LINUX}-system. The test-code
produces the test-output for the structure-functions {\tt xg1p, xg1n, xg1d} and their
1$\sigma$ errors  {\tt dxg1p, dxg1n, dxg1d}~:

\begin{verbatim}
* x,q2,g1p,dg1p,g1n,dg1n,g1d,dg1d
  0.100000  4.000000  0.027274  0.001453 -0.011366  0.001468  0.007358  0.000955
  0.200000  4.000000  0.043553  0.001347 -0.007264  0.001895  0.016784  0.001075
  0.300000  4.000000  0.052548  0.001146 -0.004617  0.001947  0.022168  0.001045
  0.400000  4.000000  0.054270  0.001137  0.000329  0.002212  0.025252  0.001150
  0.500000  4.000000  0.046195  0.001369  0.003331  0.002272  0.022906  0.001227
  0.600000  4.000000  0.034999  0.001821  0.003411  0.001920  0.017764  0.001224
  0.700000  4.000000  0.022119  0.001858  0.002789  0.001254  0.011520  0.001037
  0.800000  4.000000  0.011919  0.001254  0.001508  0.000557  0.006210  0.000634
  0.900000  4.000000  0.006516  0.000383  0.000476  0.000115  0.003234  0.000185
  0.950000  4.000000  0.005713  0.000087  0.000252  0.000023  0.002759  0.000042
\end{verbatim}

\noindent
The program can be received on request via e-mail to
{\tt Johannes.Bluemlein@desy.de} or {\tt Helmut.Boettcher@desy.de} or
from {\tt http://www-zeuthen.desy.de/\~{}blumlein}.

\newpage
\section{Tables}
\label{sec:tab}
%
\renewcommand{\arraystretch}{1.3}
%
\begin{center}
\begin{tabular}{|l|c|c|r|r|c|}
\hline \hline 
Experiment & $x$--range & $Q^2$--range  & \multicolumn{2}{c|}{data
  points} & ${\cal{N}}_i$ \\ \cline{4-5}
           &            & [$GeV^2$]     & type & \# & \\
\hline \hline
E143(p)\cite{E143pd}   & 0.027 -- 0.749 & 1.17 --  9.52 & $g_1/F_1$ &  82 & 
0.963 \\ 
HERMES(p)\cite{HERMpd} & 0.026 -- 0.731 & 1.12 -- 14.29 & $A_1$     &  37 &
0.970 \\ 
E155(p)\cite{E155p}    & 0.015 -- 0.750 & 1.22 -- 34.72 & $g_1/F_1$ &  24 &
1.003 \\ 
SMC(p)\cite{SMCpd}     & 0.004 -- 0.484 & 1.14 -- 72.10 & $A_1$     &  59 &
0.960 \\ 
EMC(p)\cite{EMCp}      & 0.015 -- 0.466 & 3.50 -- 29.5  & $A_1$     &  10 &
0.964 \\ 
CLAS1(p)\cite{CLA1pd}  & 0.125 -- 0.575 & 1.10 --  4.16 & $A_1$     &  10 &
1.010 \\
CLAS2(p)\cite{CLA2pd}  & 0.292 -- 0.592 & 1.01 --  4.96 & $g_1/F_1$ & 191 &
1.030 \\ 
COMPASS(p)\cite{COMP1} & 0.005 -- 0.568 & 1.10 -- 62.10 & $A_1$     &  15 & 
0.955 \\
\hline
proton        &                &               &           & 428 & \\
\hline \hline
E143(d)\cite{E143pd}   & 0.027 -- 0.749 & 1.17 --  9.52 & $g_1/F_1$ & 82 &
0.960 \\ 
HERMES(d)\cite{HERMpd} & 0.026 -- 0.731 & 1.12 -- 14.29 & $A_1$     &  37 &
0.970 \\
E155(d)\cite{E155d}    & 0.015 -- 0.750 & 1.22 -- 34.79 & $g_1/F_1$ &  24 &
0.979 \\ 
SMC(d)\cite{SMCpd}     & 0.004 -- 0.483 & 1.14 -- 71.76 & $A_1$     &  65 &
0.998 \\ 
COMPASS(d)\cite{COMPd} & 0.005 -- 0.566 & 1.10 -- 55.30 & $A_1$     &  15 &
0.952 \\ 
CLAS1(d)\cite{CLA1pd}  & 0.125 -- 0.575 & 1.01 --  4.16 & $A_1$     &  10 &
1.003 \\ 
CLAS2(d)\cite{CLA2pd}  & 0.298 -- 0.636 & 1.01 --  4.16 & $g_1/F_1$ & 662 &
1.014 \\ 
\hline
deuteron      &                &               &           & 895 & \\
\hline \hline
E142(n)\cite{E142n}    & 0.035 -- 0.466 & 1.10 --  5.50 & $A_1$     &  33 &
0.989 \\ 
HERMES(n)\cite{HERMn}  & 0.033 -- 0.464 & 1.22 --  5.25 & $g_1$     &   9 &
0.970 \\ 
E154(n) \cite{E154n}/\cite{E154QCD} & 0.017 -- 0.564 & 1.20 -- 15.00 & $g_1$     &  17 &
0.980 \\ 
JLAB(n) \cite{JLABn}   & 0.330 -- 0.600 & 2.71 --  4.83 & $g_1$     &   3 &
1.000 \\ 
\hline
neutron       &                &               &           &  62 & \\
\hline \hline
total         &                &               &           &1385 & \\  
\hline \hline            
\end{tabular}
\end{center}
\normalsize
\vspace{2mm}
\noindent
{\sf Table~1: Number of data points on $A_1$, $g_1/F_1$
  or $g_1$ for $Q^2 > 1.0~\GeV^2$ and $W^2 > 3.24~\GeV^2$ used in the present 
QCD analysis. For each experiment are given the $x$ and $Q^2$ ranges, 
the type of quantity measured, the number of data points for each 
given target, and the fitted normalization shifts ${\cal{N}}_i$ (see
text).}
\renewcommand{\arraystretch}{1}
%
\newpage
\renewcommand{\arraystretch}{1.3}
\begin{center}
\begin{tabular}{|c|c|c|c|c|c|}
\hline \hline
$\Delta u_v$  & $\eta$   &  0.928 (fixed)     & $\Delta \bar{q}_s$ &
                $\eta$   & -0.066 $\pm$ 0.013 \\  
              & $a$      &  0.239 $\pm$ 0.027 &                   & 
                $a$      &  0.365 $\pm$ 0.164 \\ 
              & $b$      &  3.031 $\pm$ 0.178 &                   & 
                $b$      &  8.080 (fixed)     \\ 
              & $\gamma$ & 27.64 (fixed)      &                   & 
                $\gamma$ &  0.0 (fixed)      \\ 
\hline
$\Delta d_v$  & $\eta$   & -0.342 (fixed)     & $\Delta G       $ & 
                $\eta$   &  0.462 $\pm$ 0.430 \\ 
              & $a$      &  0.128 $\pm$ 0.068 &                   & 
                $a$      &  $a_{\Delta \bar{q}_s} + 1$ \\ 
              & $b$      &  4.055 $\pm$ 0.879 &                   &  
                $b$      &  5.610 (fixed)     \\ 
              & $\gamma$ & 44.26 (fixed)      &                   & 
                $\gamma$ &  0.0 (fixed)       \\ 
\hline
\multicolumn{3}{|c|}{$\Lambda_{QCD}^{(4)} = 243 \pm 62~{\rm MeV}$} & 
\multicolumn{3}{c|}{$\chi^2 / NDF = 1537/1377 = 1.12$} \\   
\hline \hline
\end{tabular}
\end{center}
{\sf Table~2: Final parameter values and their statistical errors at
the input scale $Q_0^2=4.0~\GeV^2$. 
\renewcommand{\arraystretch}{1.0}

\vspace*{2cm}
\begin{center}
%
\renewcommand{\arraystretch}{1.3}
\footnotesize
\begin{center}
\begin{tabular}{||c||c|c|c|c|c|c|c|c||}
\hline \hline 
        & $\Lambda_{QCD}^{(4)}$ & $a_{u_v}$ & $b_{u_v}$ & $a_{d_v}$ & $ b_{d_v}$ & $\eta_{sea}$ & $a_{sea}$ & $\eta_G$ \\
\hline \hline
 $\Lambda_{QCD}^{(4)}$ & {\bf 3.84E-3} &  &  &  &  &  &  & \\
\hline
 $a_{u_v}$    & -4.08E-4 & {\bf 7.56E-4} &  &  &  &  &  & \\
\hline
 $b_{u_v}$    & -1.14E-3 &  4.30E-3 & {\bf 3.18E-2} &  &  &  &  & \\
\hline
 $a_{d_v}$    &  2.74E-3 & -9.39E-4 & -4.43E-3 & {\bf 4.60E-3} &  &  &  & \\
\hline
 $b_{d_v}$    &  2.38E-2 & -8.33E-3 & -1.03E-2 &  4.50E-2 & {\bf 7.72E-1} &  &  & \\
\hline
 $\eta_{sea}$ &  1.70E-3 & -6.84E-4 & -3.60E-3 &  2.26E-3 &  2.11E-2 &
 {\bf 5.62E-3} &  & \\
\hline
 $a_{sea}$    & -5.64E-3 &  3.04E-3 &  1.65E-2 & -8.25E-3 & -7.37E-2 &
  7.13E-4 & {\bf 2.70E-2} & \\
\hline
 $\eta_G$     & -1.96E-2 &  8.32E-3 &  4.25E-2 & -2.16E-2 & -1.67E-1 & -2.09E-2 &  4.17E-2 & {\bf 1.85E-1} \\
\hline
\hline
\end{tabular} 
\end{center}
\normalsize
\vspace{+2mm}
\noindent
{\sf Table~3: The covariance matrix for the $7+1$ parameter NLO fit
  based on the world asymmetry data.} 
\renewcommand{\arraystretch}{1.0}
\end{center}
%
\newpage
%
\renewcommand{\arraystretch}{1.1}
\begin{center}
\begin{tabular}{|l|r|r|c||r|}
\hline \hline 
\multicolumn{1}{|l|}{ }&
\multicolumn{1}{c|}{ }&
\multicolumn{2}{c||}{\sf Fit Results }  &     
\multicolumn{1}{c|}{ } \\
\cline{2-4}
\multicolumn{1}{|c|}{\sf Distribution }& 
\multicolumn{1}{ c|}{$n$} &
\multicolumn{1}{ c|}{value} &
\multicolumn{1}{ c||}{value out} & 
\multicolumn{1}{ c|}{\cite{BB}, set 3} 
\\
\multicolumn{1}{|c|}{         } &
\multicolumn{1}{ c|}{         } &
\multicolumn{1}{ c|}{         } &
\multicolumn{1}{ c||}{of range } &
\multicolumn{1}{ c|}{         } 
\\
\hline\hline
$\Delta u_v$  &  0 & $0.928 \pm 0.000$
                   & { $0.158|3.3\EM3$}
                   & {$0.926 \pm 0.071$}
\\
              &  1 & $0.153 \pm 0.004$
                   & { $1.6\EM4|2.7\EM3$}
                   & {$0.163 \pm 0.014$}
\\
              &  2 & $0.052 \pm 0.002$
                   & { $0|2.1\EM3$}
                   & $0.055 \pm 0.006$ 
\\
              &  3 & $0.023 \pm 0.001$
                   & { $0|1.7\EM3$}
                   & $0.024 \pm 0.003$
\\
\hline
$\Delta d_v$  &  0 
                   & $-0.342 \pm 0.000$
                   & { $-0.110|$$-2.1\EM4$}
                   & $-0.341 \pm 0.123$
\\
              &  1 & $-0.037 \pm 0.007$
                   & { $-7.0\EM5|$$-1.7\EM4$}
                   & $-0047 \pm 0.021$ 
\\
              &  2 & $-0.010 \pm 0.002$
                   & { $0|$ $-1.3\EM4$}
                   & $-0.015 \pm 0.009$
\\
              &  3 & $-0.004 \pm 0.001$
                   & { $0|$$-1.1\EM4$}
                   & $-0.006 \pm 0.005$
\\
\hline
$\Delta u$--$\Delta d$
              &  0 & $1.270  \pm 0.000$
                   & {  $0.267|3.5\EM3$}
                   & $1.267 \pm 0.142$
\\
              &  1 & $0.190 \pm 0.008$
                   & {  $2.3\EM4|2.8\EM3$}
                   & $0.210 \pm 0.025$
\\
              &  2 & $0.063 \pm 0.004 $
                   & { $0|2.3\EM3$}
                   & $0.070 \pm 0.011$ 
\\
              &  3 & $0.027 \pm 0.002 $
                   & { $0|1.8\EM3$}
                   & $0.030 \pm 0.006$
\\
\hline
$\Delta u$
              &  0 & $0.866 \pm 2\EM5 $
                   & {  $0.136|3.3\EM3$}
                   & $0.851 \pm 0.075$
\\
              &  1 & $0.151 \pm 0.004 $
                   & {  $1.3\EM4|2.7\EM3$}
                   &$0.160 \pm 0.014$
\\
              &  2 & $0.052 \pm 0.002 $
                   & { $0|2.1\EM3$}
                   & $0.055 \pm 0.006$
\\
              &  3 & $0.023 \pm 0.001 $
                   & { $0|1.7\EM3$}
                   & $0.024 \pm 0.003$
\\
\hline
$\Delta d$
              &  0 & $-0.404 \pm 3\EM5 $
                   & {  $-0.132|$$-2.1\EM4$}  
                   & $-0.415 \pm 0.124 $
\\
              &  1 & $-0.039 \pm 0.007 $
                   & { $-1.0\EM4|$$-1.7\EM4$}
                   & $-0.050 \pm 0.022$
\\
              &  2 & $-0.011 \pm 0.002 $
                   & { -$0|$$-1.3\EM4$}
                   & $-0.015 \pm 0.009$
\\
              &  3 & $-0.004 \pm 0.001 $
                   & { $0|$$-1.1\EM4$}
                   & $-0.006 \pm 0.005$
\\
\hline
$\Delta \overline{q}$
              &  0 & $-0.066 \pm 0.013 $
                   & { $-0.02|0$}
                   & $-0.074 \pm 0.017$
\\
              &  1 & $-2.5\EM3 \pm 1.2\EM3 $
                   & { $-3.0\EM5|0$}
                   & $-0.003 \pm 0.001$
\\
              &  2 & $-3.3\EM4 \pm 2.0\EM4$
                   & {$0|0$}
                   & $-4.0\EM4 \pm 1.0\EM4$
\\
              &  3 & $-7.0\EM5 \pm 4.0\EM5$
                   & {$0|0$}
                   & $-8.0\EM5 \pm 2.0\EM5$ 
\\
\hline
$\Delta G$    &  0 & $ 0.462 \pm 0.430 $
                   & { $0.004|1.0\EM4$}
                   & $1.062 \pm 0.549$
\\
              &  1 & $ 0.079 \pm 0.079 $
                   & {$1.0\EM5|8.0\EM5$}
                   & $ 0.184 \pm 0.103$
\\
              &  2 & $ 0.021 \pm 0.021 $
                   & {$0|6.3\EM5$}
                   & $0.050 \pm 0.028$ 
\\
              &  3 & $ 0.007 \pm 0.007 $
                   & {$0|4.9\EM5$}
                   & $0.017 \pm 0.010$
\\
\hline
\hline
\end{tabular}
\end{center}
\normalsize

\vspace{2mm}
\noindent
{\sf Table~4: Moments of the NLO parton densities and their
combinations for the present analysis at $Q^2 = 4~\GeV^2$. The value
of the respective moment integrating only outside the $x$--range in which
currently deep--inelastic scattering data are measured, $0.005 < x < 0.75$,
are given for comparison (lower$|$upper~part). The errors are the 
$1\sigma$ correlated errors. 
\renewcommand{\arraystretch}{1}
%
\newpage
\renewcommand{\arraystretch}{1.3}
\begin{center}
\begin{tabular}{|c|c|c|c|c|}
\hline \hline
$<x>$  & $C_p {\rm [GeV^2]}$ & $C_d {[\rm GeV^2]}$ &
         $C_p {\rm [GeV^2]}$ & $C_d {[\rm GeV^2]}$ \\
\cline{2-5}
& \multicolumn{2}{c|}{multiplicative }& \multicolumn{2}{c|}{additive } \\
\hline
0.060  & -0.084 $\pm$ 0.245     &  0.007 $\pm$ 0.287 
       &  0.011 $\pm$ 0.076     &  0.019 $\pm$ 0.059 \\
0.150  & -0.229 $\pm$ 0.156     &  0.169 $\pm$ 0.431 
       & -0.038 $\pm$ 0.037     &  0.022 $\pm$ 0.042 \\
0.275  & -0.224 $\pm$ 0.099     & -0.226 $\pm$ 0.270 
       & -0.045 $\pm$ 0.022     & -0.007 $\pm$ 0.025 \\
0.425  & -0.083 $\pm$ 0.140     & -0.013 $\pm$ 0.384 
       & -0.017 $\pm$ 0.021     &  0.005 $\pm$ 0.027 \\
0.625  &  0.290 $\pm$ 0.417     &  0.061 $\pm$ 1.106 
       &  0.011 $\pm$ 0.033     &  0.011 $\pm$ 0.031 \\
\hline \hline
\end{tabular}
\end{center}
\normalsize
\vspace{2mm}
\noindent
{\sf Table~5: The higher twist coefficients $C_p(x)$ and $C_d(x)$
  as function of $x$ considering both a multiplicative and additive phenomenological ansatz.}
\renewcommand{\arraystretch}{1}
%
%
\newpage
\section{Figures}
\label{sec:fig}
\begin{figure}[htb]
\begin{center}
\includegraphics[angle=0, width=14.0cm]{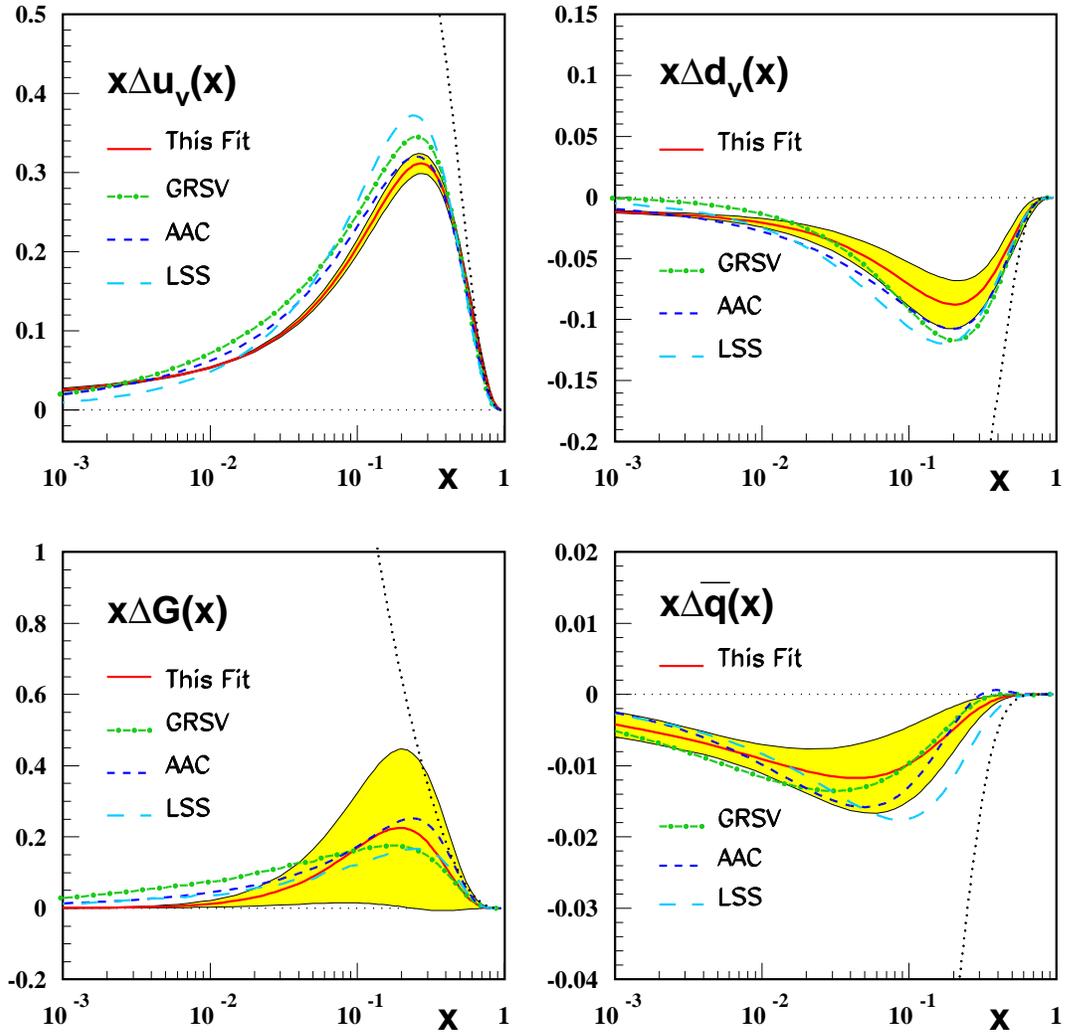}
\end{center}
\caption{\label{fig:xpdf}
\sf
NLO polarized parton distributions at the input scale 
$Q_0^2 = 4.0~\GeV^2$ (solid line) compared to results obtained by 
GRSV~(dashed--dotted line)~\cite{GRSV}, DSSV~(long dashed--dotted
line)~\cite{DSSV}, AAC~(dashed line)~\cite{AAC}, and LSS~(long dashed
line)~\cite{LSS}. 
The shaded areas represent the fully correlated $1\sigma$ error bands 
calculated by Gaussian error propagation.} 
\end{figure}

\newpage

\vspace*{2cm}
\begin{figure}[htb]
\begin{center}
\includegraphics[angle=0, width=14.0cm]{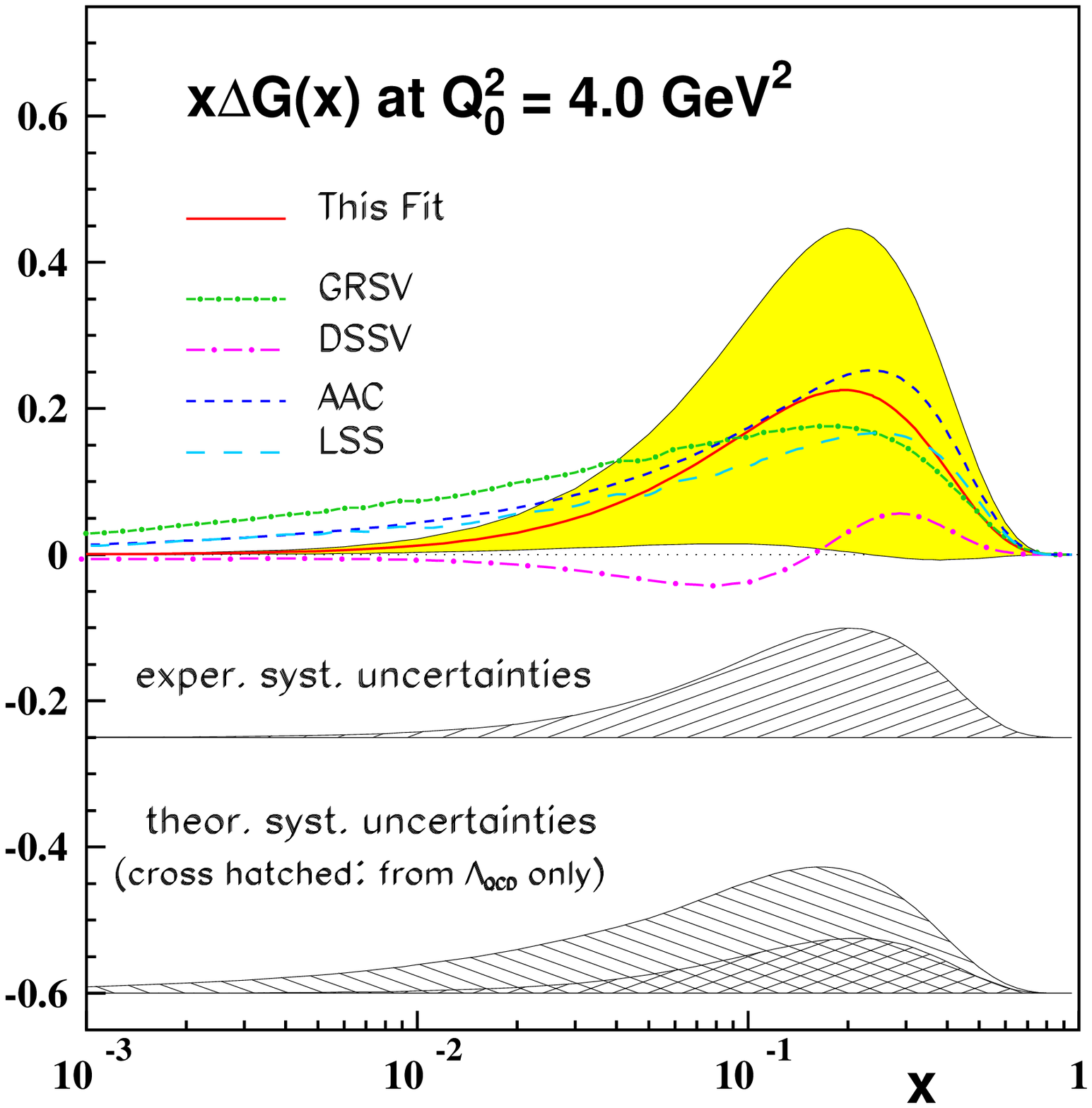}
\end{center}
\caption{\label{fig:xgl}
\sf
The polarized parton density $x\Delta G(x)$ at $Q_0^2 = 4.0~$GeV$^2$ 
as a function of $x$ (solid line). The shaded area is the fully
correlated $1\sigma$ statistical error band and the hatched areas are
the systematic uncertainties. Results from GRSV~(dashed--dotted
line)~\cite{GRSV}, DSSV~(long dashed--dotted line)~\cite{DSSV},  
AAC~(dashed line)~\cite{AAC}, and LSS~(long dashed line)~\cite{LSS}
are shown for comparison.} 
\end{figure}

\newpage

\vspace*{2cm}
\begin{figure}[htb]
\begin{center}
\includegraphics[angle=0, width=14.0cm]{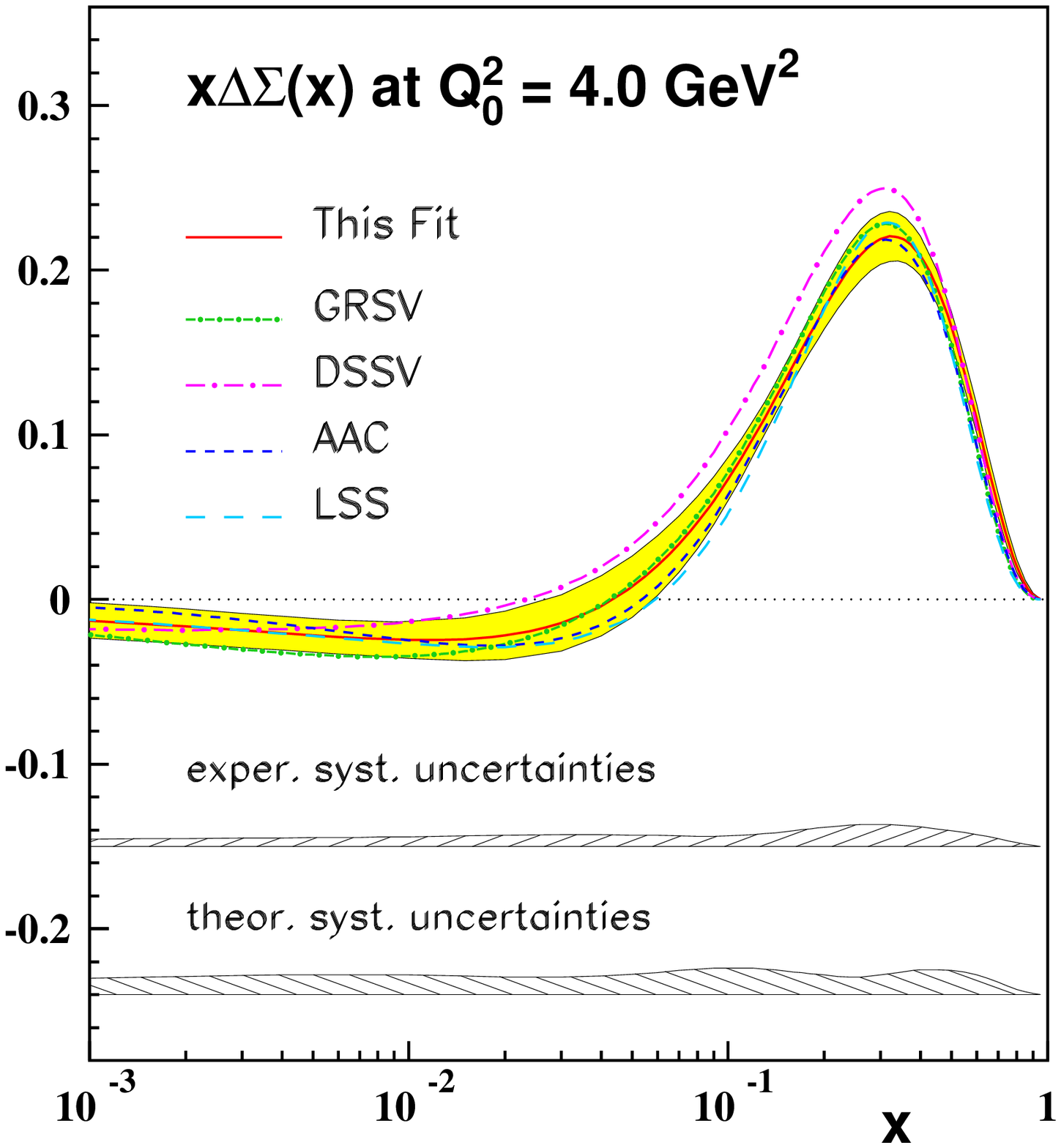}
\end{center}
\caption{\label{fig:xsng}
\sf
The polarized parton density $x\Delta \Sigma(x)$ at $Q_0^2 =
4.0~$GeV$^2$ as a function of $x$ (solid line). The shaded area is the
fully correlated $1\sigma$ statistical error band and the hatched
areas are the systematic uncertainties. 
Results from GRSV~(dashed--dotted line)~\cite{GRSV}, DSSV~(long
dashed--dotted line)~\cite{DSSV}, AAC~(dashed line)~\cite{AAC}, and 
LSS~(long dashed line)~\cite{LSS} are shown for comparison.} 
\end{figure}

\vspace*{-2.0cm}
\begin{figure}[htb]
\begin{center}
\includegraphics[angle=0, width=11.0cm]{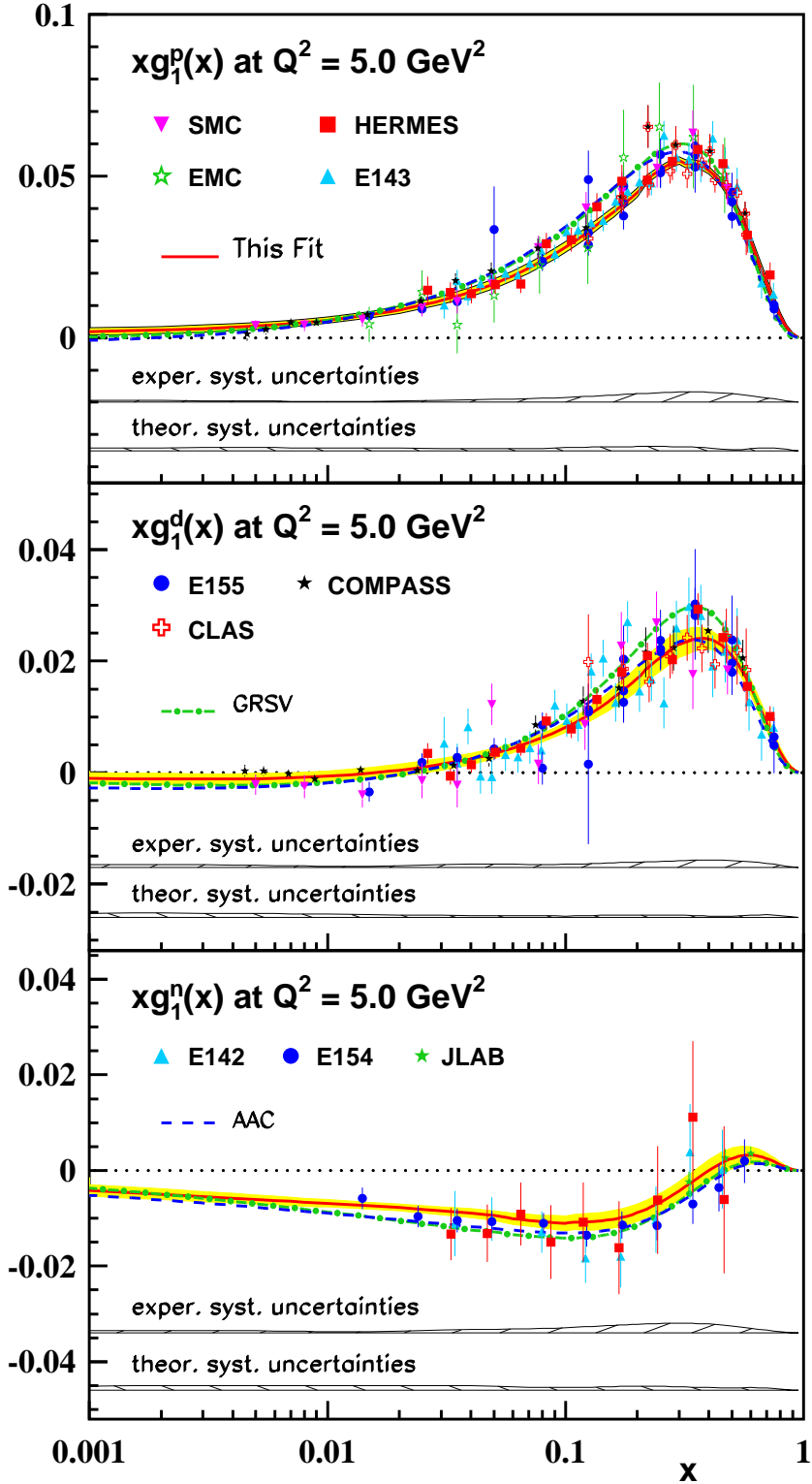}
\end{center}
\caption{\label{fig:g1pdn}
\sf
The spin--dependent structure functions $xg_1^p(x)$, $xg_1^d(x)$ and
$xg_1^n(x)$ as a function of $x$. 
The experimental data are evolved to a common value of $Q^2 = 5~{\rm
GeV^2}$. The error bars shown are the statistical and systematic ones
added in quadrature. The experimental distributions are well described
(solid curve) within the statistical (shaded areas) and systematic
(hatched areas) error bands. The curves obtained by GRSV~(dashed-dotted)~\cite{GRSV} 
and AAC (dashed)~\cite{AAC} are shown for comparison.}
\end{figure}

\vspace*{-2.0cm}
\begin{figure}[htb]
\begin{center}
\includegraphics[angle=0, width=14.0cm]{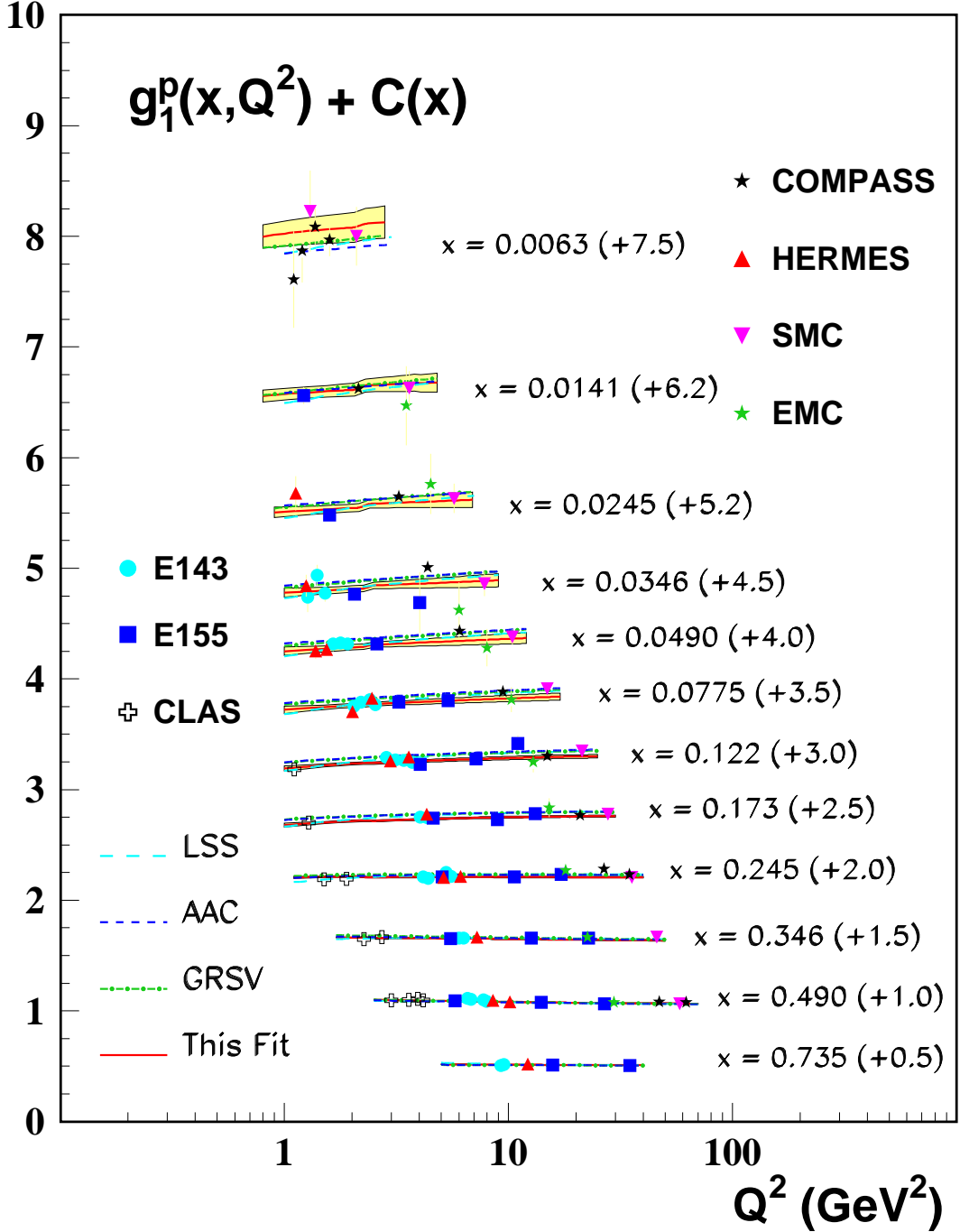}
\end{center}
\caption{\label{fig:g1pdn1}
\sf
The spin--dependent structure function $xg_1^p(x,Q^2)$  as a function of $x$ and $Q^2$.
The experimental data are compared to the fit result (solid curve)
with the statistical error bands (shaded areas). The curves
obtained by GRSV~(dashed-dotted)~\cite{GRSV}, AAC (short dashed)~\cite{AAC} 
and LSS (long dashed)~\cite{LSS} are shown for comparison.}
\end{figure}

\begin{figure}[htb]
\begin{center}
\includegraphics[angle=0, width=14.0cm]{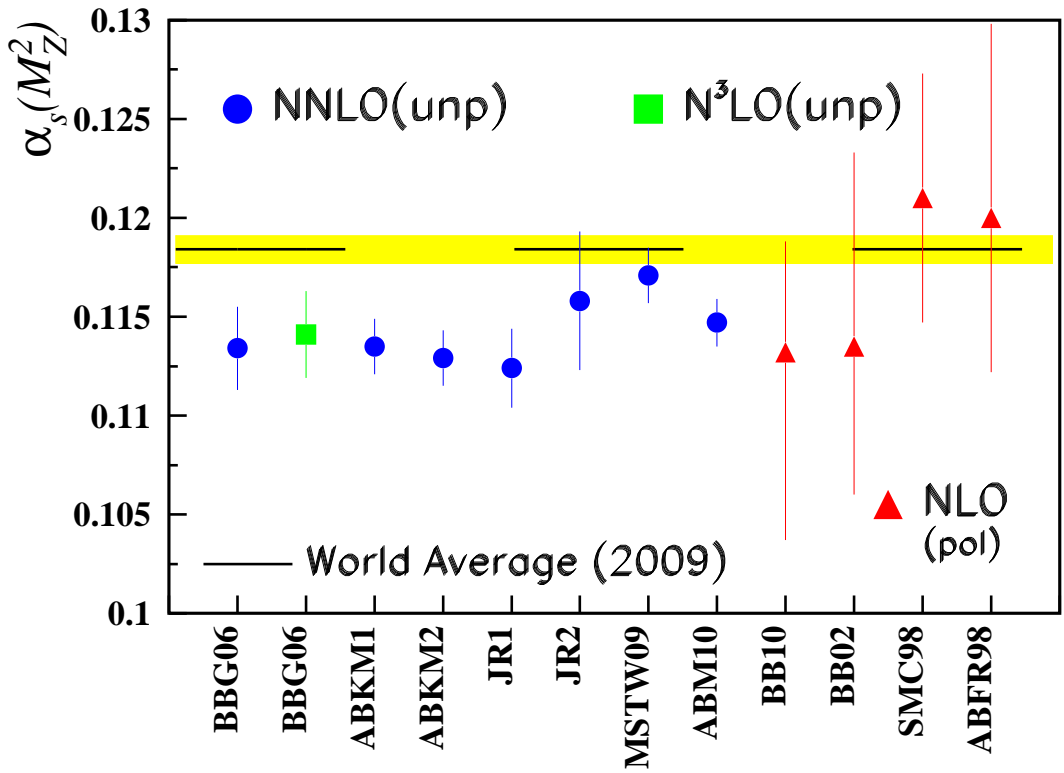}
\end{center}
\caption{\label{fig:alps}
\sf
The strong coupling constant $\alpha_s(M_Z^2)$ from different DIS
measurements, Eqs.~(49--56) and Refs.~[22,25,17]. The yellow band describes
the weighted average of a wide range of $\alpha_s(M_Z)$ measurements \cite{BETHKE}.}
\end{figure}
\begin{figure}[htb]
\begin{center}
\includegraphics[angle=0, width=14.0cm]{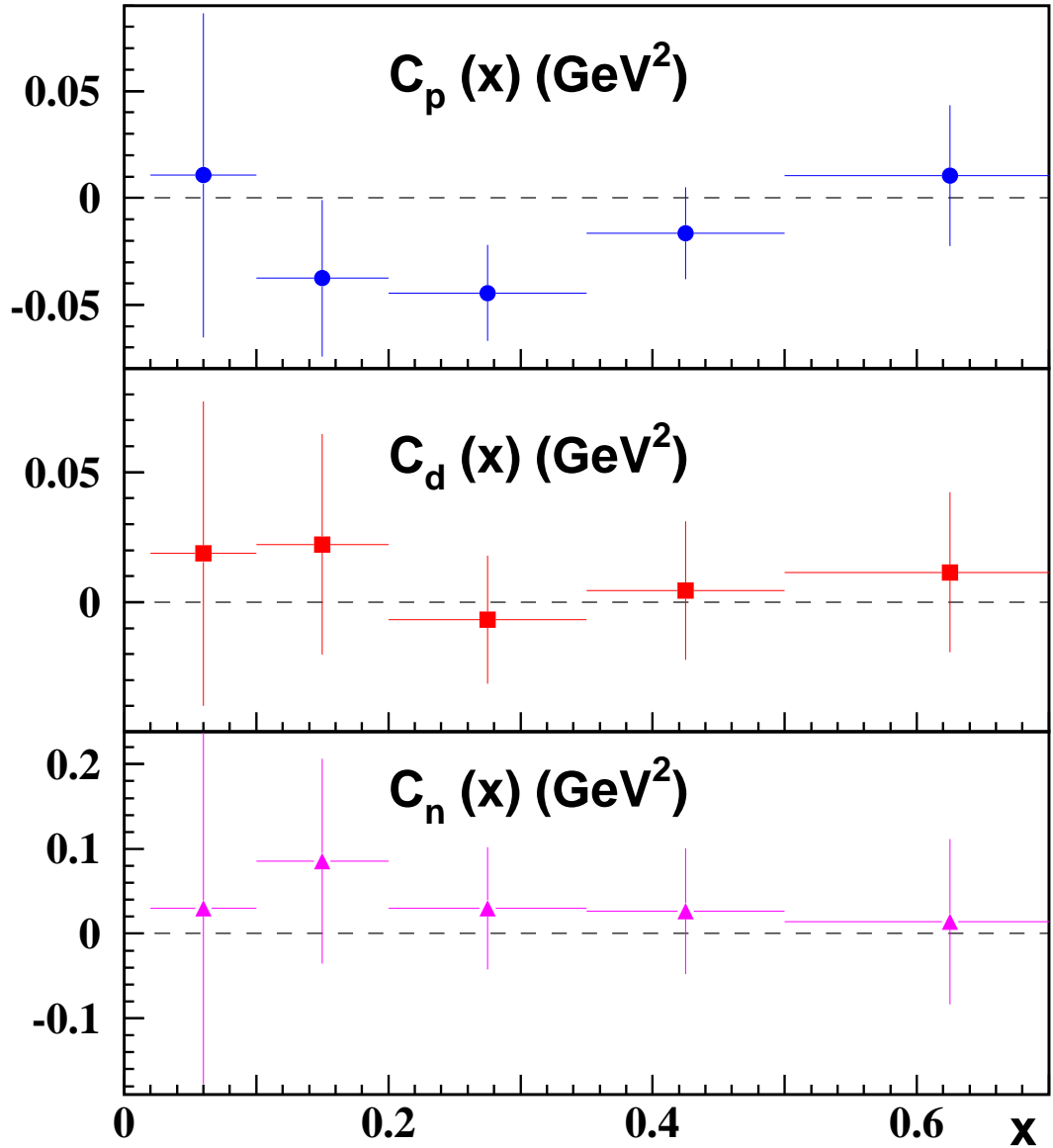}
\end{center}
\caption{\label{fig:ht}
\sf
The additive higher twist coefficients $C_p(x)$, $C_d(x)$ and $C_n(x)$ as
a function of $x$.}
\end{figure}
%
\clearpage
\rm

\end{document}